\documentclass[prd,aps,twocolumn,preprintnumbers, showpacs, nofootinbib,superscriptaddress,notitlepage]{revtex4-1}
\usepackage{amssymb,amsthm,amsmath}
\usepackage{graphicx}   % figures
\usepackage{color}      % color is used in text
\usepackage{slashed}    % Feynman slash
\usepackage[normalem]{ulem}

\usepackage{rotating}   % to rotate tables
\usepackage{multirow}   % multicolumn and multirow
\usepackage[normalem]{ulem}

\newcommand{\nn}{\nonumber}

\begin{document}

\preprint{MIT-CTP/5030}

\title{Unpolarized isovector quark distribution function from Lattice QCD: \\ A systematic analysis of renormalization and matching}

\collaboration{\bf{Lattice Parton Collaboration ($\rm {\bf LPC}$)}}

\author{Yu-Sheng Liu}
%\email{mestelqure@gmail.com}
\affiliation{Tsung-Dao Lee Institute, Shanghai Jiao Tong University, Shanghai 200240, China}

\author{Jiunn-Wei Chen}
%\email{jwc@phys.ntu.edu.tw}
\affiliation{Department of Physics, Center for Theoretical Physics, and Leung Center for Cosmology and Particle Astrophysics, National Taiwan University, Taipei, Taiwan 106}
\affiliation{Center for Theoretical Physics, Massachusetts Institute of Technology, Cambridge, MA 02139, USA}

\author{Yi-Kai Huo}
\affiliation{INPAC, SKLPPC, MOE KLPPC, School of Physics and Astronomy, Shanghai Jiao Tong University, Shanghai 200240, China}
\affiliation{Zhiyuan College, Shanghai Jiao Tong University, Shanghai 200240, China}

\author{Luchang Jin}
%\email{ljin.luchang@gmail.com}
\affiliation{Physics Department, University of Connecticut, Storrs, Connecticut 06269-3046, USA}
\affiliation{RIKEN BNL Research Center, Brookhaven National Laboratory, Upton, NY 11973, USA}

\author{Maximilian Schlemmer}
\affiliation{Institut f\"ur Theoretische Physik, Universit\"at Regensburg, D-93040 Regensburg, Germany}

\author{Andreas Sch\"afer}
\affiliation{Institut f\"ur Theoretische Physik, Universit\"at Regensburg, D-93040 Regensburg, Germany}

\author{Peng Sun}
\email{Corresponding author: 06260@njnu.edu.cn}
\affiliation{Nanjing Normal University, Nanjing, Jiangsu, 210023, China}

\author{Wei Wang}
\email{Corresponding author: wei.wang@sjtu.edu.cn}
\affiliation{INPAC, SKLPPC, MOE KLPPC, School of Physics and Astronomy, Shanghai Jiao Tong University, Shanghai 200240, China}

\author{Yi-Bo Yang}
\affiliation{Department of Physics and Astronomy, Michigan State University, East Lansing, MI 48824, USA}
\affiliation{CAS Key Laboratory of Theoretical Physics, Institute of Theoretical Physics, Chinese Academy of Sciences, Beijing 100190, China}

\author{Jian-Hui Zhang}
\affiliation{Institut f\"ur Theoretische Physik, Universit\"at Regensburg, D-93040 Regensburg, Germany}
\affiliation{Center of Advanced Quantum Studies, Department of Physics, Beijing Normal University, Beijing 100875, China}

\author{Qi-An Zhang}
\affiliation{Tsung-Dao Lee Institute, Shanghai Jiao Tong University, Shanghai 200240, China}

\author{Kuan Zhang}
\affiliation{University of Chinese Academy of Sciences, School of Physical Sciences, Beijing 100049, China}
\affiliation{CAS Key Laboratory of Theoretical Physics, Institute of Theoretical Physics, Chinese Academy of Sciences, Beijing 100190, China}

\author{Yong Zhao}
\affiliation{Center for Theoretical Physics, Massachusetts Institute of Technology, Cambridge, MA 02139, USA}
\affiliation{Physics Department, Brookhaven National Laboratory,  {Bldg. 510A,} Upton, NY 11973, USA}

\date{\today}

\begin{abstract}
We present a detailed Lattice QCD study of the unpolarized isovector quark Parton Distribution Function (PDF) using   large-momentum effective theory framework. We choose  a quasi-PDF  defined by a spatial correlator which is free from mixing with other operators of the same dimension.  In the lattice simulation, we use a Gaussian-momentum-smeared source  at $M_\pi=356$~MeV and $P_z \in \{1.8,2.3\}$~GeV.  To control the systematics associated with the excited states, we explore  {five different source-sink separations}.  The nonperturbative renormalization is conducted in a regularization-independent momentum subtraction scheme, and the matching between the renormalized quasi-PDF and $\overline{\rm MS}$ PDF is calculated based on perturbative QCD up to one-loop order. Systematic errors due to renormalization and perturbative matching are also analyzed in detail.  Our results for lightcone PDF are in reasonable agreement with the latest phenomenological analysis.
\end{abstract}
\maketitle

\section{Introduction}
Parton distribution functions (PDFs) of nucleons are not only  important quantities characterizing   the internal hadron  structures but  are also key ingredients to make   predictions for high-energy scattering processes~\cite{Butterworth:2015oua,Alekhin:2017kpj,Gao:2017yyd}. Thus calculating PDFs from first principles has been a holy grail in nuclear and particle physics. Since PDFs are embedded with the low-energy quark and gluon degrees of freedom in the hadron, they involve infrared (IR) dynamics of strong interactions  and can only be determined by nonperturbative methods such as Lattice QCD.

Within QCD factorization~\cite{Collins:2011zzd}, the quark PDF is defined  as
\begin{align} \label{eq:lcpdf}
q(x,\mu) \equiv\! \int\! {d\xi^-\over 4\pi} e^{-ixP^+\xi^-}\! \langle P | \bar{\psi}  (\xi^-) \gamma^+ U (\xi^-, 0) \psi (0) | P \rangle ,
\end{align}
where $|P\rangle$ denotes the nucleon state with momentum $P_\mu=(P_t,0,0,P_z)$. $x$ is the quark momentum fraction, $\mu$ is the renormalization scale in the $\overline{\rm MS}$ scheme.  $\xi^{\pm}=(t\pm z)/\sqrt{2}$ are  the lightcone coordinates. The light-like Wilson line is introduced to maintain the gauge invariance:
\begin{align}
U(\xi^-,0) = P\exp\biggl(-ig\int_0^{\xi^-} d\eta^- A^+(\eta^-)\biggr)\ .
\end{align}

PDFs are defined with  lightcone coordinates, but  the Lattice simulation can only be conducted in  Euclidean space with no proper treatment for lightcone quantities which involves real time. Thus
simulating PDFs on a Euclidean Lattice is an extremely difficult task.
Early studies based on operator product expansion (OPE) were only able to derive the lowest few moments of the PDFs~\cite{Martinelli:1987zd,Martinelli:1988xs,Detmold:2001dv,Dolgov:2002zm}.

Recently, a novel approach  that allows  to directly access the $x$-dependence of PDFs  from Lattice QCD was proposed in Ref.~\cite{Ji:2013dva}, now formulated as large-momentum effective theory (LaMET)~\cite{Ji:2014gla}. Within this framework, one can  extract PDFs---as well as other lightcone quantities---from the correlations of certain static operators in a  nucleon state. On the one hand,  the  static correlations, often referred to as quasi observables, can be directly calculated on a Euclidean Lattice and depend dynamically on the nucleon momentum. On the other hand, at large momentum,  the quasi observables can be factorized into the parton observable and a perturbative matching coefficient, up to corrections suppressed by    powers of the large nucleon momentum. Equating the results from the two sides provides a straightforward way to determine the lightcone PDFs.

To calculate the quark PDF in LaMET, one starts with a ``quasi-PDF" which is defined as an equal-time correlation of quarks along the $z$ direction~\cite{Ji:2013dva}: 
\begin{align}\label{eq:qpdf}
\widetilde{q}_{\Gamma}(x,P_z) \equiv \int_{-\infty}^\infty {dz\over 4\pi} e^{ixP_zz}\langle P |O_{\Gamma}(z)| P \rangle \,.
\end{align}
In the above,
$O_{\Gamma}(z)=\bar{\psi}  (z) \Gamma U (z, 0) \psi (0)$ with $\Gamma=\gamma^z$ or $\Gamma=\gamma^t$, and the space-like Wilson line  is:
\begin{align}
U(z,0) = P\exp\left(-ig\int_0^z dz' A_z(z')\right) .
\end{align}
For finite but large momentum $P_z$, $\widetilde{q}(x,P_z)$ has support in $-\infty < x < \infty$.
 Unlike the lightcone PDF that is boost invariant, the quasi-PDF has a nontrivial dependence on the nucleon momentum $P_z$. 
After renormalizing the quasi-PDF  in a scheme  such as the regularization-independent momentum subtraction (RI/MOM) scheme, one can match the renormalized quasi-PDF to the $\overline{\rm MS}$ PDF  through   the factorization theorem~\cite{Ji:2013dva,Ji:2014gla,Stewart:2017tvs,Izubuchi:2018srq,Ma:2014jla,Ma:2017pxb}: 
\begin{align} \label{eq:fact}
\widetilde{q}(x,P_z, p^R_z,\mu_R)=&\int_{-1}^1 {dy\over |y|}\: C\left({x\over y},r,\frac{yP_z}{\mu},\frac{yP_z}{p_z^R}\right) \, q(y,\mu)\nonumber\\
&+\mathcal{O}\left({M^2\over P_z^2},{\Lambda_{\text{QCD}}^2\over x^2 P_z^2}\right) ,
\end{align}
where $p_z^R$ and $\mu_R$ are introduced in RI/MOM scheme: $p_z^R$  is  the momentum of the involved  parton and $\mu_R$ is renormalization scale.  $r=\mu_R^2/(p_z^R)^2$, $C$ is the perturbative matching coefficient, and $\mathcal{O}(M^2/P_z^2, \Lambda_{\rm QCD}^2/x^2P_z^2)$ denotes nucleon  mass and higher-twist contributions suppressed by   powers of the large nucleon momentum. The flavor indices in  $q,\ \widetilde{q}$ and   $C$ are implied. When $-1<y<0$, the distributions   refer  to  the antiquark distributions.

Since the proposal of LaMET, remarkable  progress has been made in both theoretical aspect  and    Lattice calculations. It should be pointed out that these developments are achieved  in an interactive way.   The LaMET was first used to calculate  the proton isovector quark distribution $f_{u-d}$~\cite{Lin:2014zya,Alexandrou:2015rja,Chen:2016utp,Alexandrou:2016jqi,Chen:2018fwa,Alexandrou:2018eet}, including the unpolarized, polarized and transversity cases, and subsequently  to the meson distribution amplitudes~\cite{Zhang:2017bzy,Chen:2017gck}. The first Lattice studies    used  the matching coefficients at   one-loop order in a transverse-momentum cutoff scheme~\cite{Xiong:2013bka,Ji:2015qla,Xiong:2015nua}. However, as was found in Ref.~\cite{Xiong:2013bka}, the original  quasi-PDF suffers from an ultraviolet (UV) linear divergence which might pose a severe problem for the renormalization of its Lattice matrix elements~\cite{Li:2016amo,Rossi:2017muf,Rossi:2018zkn}. Then many attentions have been paid to the renormalization property~\cite{Ji:2015jwa,Ishikawa:2016znu,Chen:2016fxx,Xiong:2017jtn,Constantinou:2017sej,Ji:2017oey,Ishikawa:2017faj,Green:2017xeu,Spanoudes:2018zya}, and finally  the  multiplicative renormalizability of quasi-PDF in coordinate space  in the continuum was proven to all orders in strong coupling constant $\alpha_s$~\cite{Ji:2017oey,Ishikawa:2017faj}. This finding has further  motivated the Lattice analysis of nonperturbative renormalization (NPR) of the quasi-PDF~\cite{Alexandrou:2017huk,Chen:2017mzz,Green:2017xeu} in the RI/MOM scheme~\cite{Martinelli:1994ty}, and the  calculation of the matching coefficients between the RI/MOM quasi-PDFs and $\overline{\rm MS}$ PDFs~\cite{Stewart:2017tvs}.
Besides the renormalization, the finite  nucleon mass corrections were also worked out to all orders of $M^2/P_z^2$~\cite{Chen:2016utp}, and  higher-twist $\mathcal O(\Lambda^2_{\rm QCD}/x^2P_z^2)$ effects were numerically removed by extrapolating the results at  {several $P_z$ values}  to infinite momentum~\cite{Lin:2014zya,Chen:2016utp}.
Based on these studies, calculations of the isovector quark PDF at physical pion mass have become available~\cite{Lin:2017ani,Alexandrou:2018pbm,Chen:2018xof,Alexandrou:2018eet}. Potential operator mixing in the Lattice renormalization of the quasi-PDF has also been investigated~\cite{Constantinou:2017sej,Alexandrou:2017huk,Green:2017xeu,Chen:2017mzz}, and  the mixing pattern classified in Ref.~\cite{Chen:2017mie}.
Ways to  {reduce} the systematic uncertainties from Fourier transforming the spatial correlation at long distance  {were}   proposed in Refs.~\cite{Lin:2017ani,Chen:2017lnm}.
The  LaMET was also attempted   to  study   transverse-momentum-dependent distributions~\cite{Ji:2014hxa,Ji:2018hvs,Ebert:2018gzl,Ebert:2019okf,Ebert:2019tvc,Ji:2019sxk,Shanahan:2019zcq,Ji:2019ewn}, as well as the gluon PDF~\cite{Wang:2017qyg,Wang:2017eel,Fan:2018dxu,Zhang:2018diq,Li:2018tpe,Wang:2019tgg}.

In addition to LaMET,
other interesting  approaches have   been proposed in recent years to calculate the PDFs from Lattice QCD. For example, one can extract the PDFs from a  class of ``Lattice cross sections"~\cite{Ma:2014jla,Ma:2017pxb}, while a smeared quasi-PDF in the gradient flow method was proposed to sweep  the power divergence in the Lattice calculation~\cite{Monahan:2016bvm,Monahan:2017hpu}.
One can also study a pseudo distribution~\cite{Radyushkin:2017cyf}, related to the quasi-PDF through Fourier transforms. While this method shows  interesting renormalization features~\cite{Orginos:2017kos,Karpie:2017bzm}, it  coincides  with  LaMET regarding the factorization
into PDFs~\cite{Ji:2017rah,Zhang:2018ggy,Izubuchi:2018srq}. Moreover  there are proposals using current-current correlators
to compute the hadronic tensor~\cite{Liu:1993cv,Liang:2017mye}, or the higher moments of the PDF, lightcone distribution amplitudes, etc.~\cite{Detmold:2005gg,Braun:2007wv,Liang:2017mye,Chambers:2017dov,Bali:2018spj,Bali:2019ecy}. These different approaches are subject to their own systematics, but they can be compared to each other.

It was argued  that the power divergent mixing between   local moment operators may spoil the renormalization of quasi-PDFs~\cite{Rossi:2017muf,Rossi:2018zkn}, however such problem dissolves in LaMET since  one  first needs to take  the continuum limit of the quasi-PDF after renormalization on the Lattice, and then match it to obtain the $x$-dependence of the PDF.  The factorization has been derived rigorously~\cite{Ma:2014jla,Izubuchi:2018srq} in the continuum, and one only needs to  focus on the renormalization of the nonlocal spatial correlator only.   Thus the renormalization of   local moment operators is irrelevant to quasi-PDF. Besides, there are also confusions on the LaMET matching between Minkowskian and Euclidean matrix elements of the quasi-PDF~\cite{Carlson:2017gpk}, which have been clarified in Refs.~\cite{Ji:2017rah,Briceno:2017cpo}.

Most of the available  Lattice calculations  have used  $\Gamma=\gamma^z$ (except \cite{Chen:2018xof,Alexandrou:2018pbm}) for the unpolarized quasi-PDF,  which is now known to mix with the scalar quasi-PDF operator  $O_{I}$ at $O(a^0)$~\cite{Constantinou:2017sej,Chen:2017mie,Green:2017xeu}.   This  operator mixing introduces an additional systematic uncertainty in   nonperturbative renormalization~\cite{Alexandrou:2017huk,Chen:2017mzz,Green:2017xeu,Lin:2017ani}, thus limiting the accuracy of the extracted PDF. On the contrary, the $\Gamma=\gamma^t$ case is free from operator mixing with $O_{I}$ at $O(a^0)$~\cite{Constantinou:2017sej,Chen:2017mie,Green:2017xeu}. Therefore, it is highly desirable to start from the quasi-PDF with $\Gamma=\gamma^t$.  This is one main motif of this study.

In this work, we will carry out a Lattice calculation of the unpolarized isovector quark distribution from the quasi-PDF with $\Gamma=\gamma^t$ with the same nonperturbative renormalization procedure as  for the $\Gamma=\gamma^z$ case in Ref.~\cite{Chen:2017mzz}.  The calculation is performed using clover fermions on a CLS ensemble of gauge configurations with $N_f=2+1$ (degenerate up/down, and strange) flavors under   open boundary condition~\cite{Luscher:2011kk} with pion mass $M_{\pi}=356$~MeV and Lattice spacing $a=0.086$~fm~\cite{Bruno:2014jqa}.  We will  examine the dependence on the nucleon momentum $P_z$ and the RI/MOM scales $p_z^R$, $\mu_R$, as well as on   choices of the projection operator for the amputated Green's function in RI/MOM renormalization.
Due to large uncertainties, it is hard  to see the sea quark asymmetry observed in early studies which were performed without Lattice renormalization~\cite{Lin:2014zya,Alexandrou:2015rja,Chen:2016utp,Alexandrou:2016jqi}.  In the future we plan to analyze CLS ensembles with a better accuracy and  down to physical masses and $a<0.04$~fm, both for $(m_s+m_u+m_d)$ fixed to its physical value and for physical $m_s$ using flavour SU(3) and SU(2) extrapolations.

The rest of this paper is organized as follows.  In Sec.~\ref{sec:matching}, we briefly  review the procedure of nonperturbative renormalization and matching of the quasi-PDF in the RI/MOM scheme, in particular  the explicit one-loop matching coefficient for the $\Gamma=\gamma^t$ case. In Sec.~\ref{sec:numerical}, we describe the details of Lattice simulation of the hadronic matrix elements as well as its nonperturbative renormalization. Systematic errors in the calculation are also discussed in this section. In Sec.~\ref{sec:final_results}, we present our   results on the $x$-dependence of the unpolarized isovector quark PDF with the statistical and systematic uncertainties, and the last section contains the summary of  our work.

\section{Nonperturbative Renormalization and Matching}\label{sec:matching}

To recover  the continuum limit of a quasi-PDF matrix element,
nonperturbative renormalization on the Lattice is required to  deal with linear and logarithmic UV divergences. In this work, we follow the RI/MOM scheme elaborated in Refs.~\cite{Stewart:2017tvs,Chen:2017mzz}, and match the result to the $\overline{\rm MS}$ PDF with the one-loop matching coefficient~\cite{Stewart:2017tvs}.

\subsection{RI/MOM renormalization on the Lattice}

The spatial correlator $O_\Gamma(z)$ has been proven to be multiplicatively renormalizable in coordinate space in the continuum~\cite{Ji:2017oey,Ishikawa:2017faj}, which enables  the renormalization in   RI/MOM scheme~\cite{Martinelli:1994ty}.

For each value of $z$, the RI/MOM renormalization factor $Z$ is obtained by requiring loop corrections for the matrix element of a quasi-PDF operator vanish in an off-shell quark state at a given momentum:
\begin{align}\label{eq:Z}
Z(z,p^R_z,a^{-1},\mu_R)=\left.\frac{\sum_s\langle p,s|O_{\gamma^t}(z)|p,s\rangle}{\sum_s\langle p,s|O_{\gamma^t}(z)|p,s\rangle_{\rm tree}}\right|_{\tiny\begin{matrix}p^2=-\mu_R^2 \\ \!\!\!\!p_z=p^R_z\end{matrix}}\,.
\end{align}
The bare matrix element $\sum_s\langle p,s|O_{\gamma^t}(z)|p,s\rangle$ will be calculated on the Lattice from the amputated Green's function $\Lambda_{\gamma^t}(p,z)$ of $O_{\gamma^t}(z)$,  with a projection operator ${\cal P}$ for the Dirac matrix:
\begin{equation}
\sum_s\langle p,s|O_{\gamma^t}(z)|p,s\rangle = \mbox{Tr}\left[ \Lambda_{\gamma^t}(z,p){\cal P}\right]\,.
\end{equation}
Due to the breaking of Lorentz covariance in $O_\Gamma(z)$, the RI/MOM subtraction  depends on  two scales $\mu_R$ and $p_z^R$. As a result, the renormalization factor $Z(z,p^R_z,a^{-1},\mu_R)$ depends on the Lattice spacing as well as on the two RI/MOM scales $\mu_R$ and $p^R_z$.

Based on the symmetry of $O_\Gamma(z)$ on the Lattice, the amputated Green's function $\Lambda_{\gamma^t}(p,z)$ is not only proportional to the tree-level result $\gamma^t$, but also includes two other independent Lorentz structures:
\begin{equation}\label{eq:ME_decomposition}
\Lambda_{\gamma^t}(p,z)=\widetilde{F}_t(p,z) \gamma^t  + \widetilde{F}_{z}(p,z)\frac{p_t\gamma^z}{p_z} +\widetilde{F}_p(p,z)\frac{p_t\slashed{p}}{p^2}\,.
\end{equation}
In the above $\widetilde{F}_i$s are independent  form factors that are invariant under the hyper cubic group $H(4)$. According to Eq. (\ref{eq:ME_decomposition}) the RI/MOM renormalization factor $Z$ will also depend on  the projection operator ${\cal P}$. One can choose to single out  $\widetilde{F}_t$ only~\cite{Stewart:2017tvs}, which we call the minimal projection. This projection  has the simplest form but captures all the UV divergence in $\Lambda_{\gamma^t}(p,z)$. Optionally, one  can choose ${\cal P} = \slashed p/(4p^t)$, which we call the $\slashed p$ projection. The renormalization factors $Z$ with the minimal and $\slashed p$ projections are defined as: 
\begin{align}\label{eq:def_renorm}
&Z_{mp}(z, p^R_z, a^{-1}, \mu_R)\equiv \widetilde{F}_t(p,z)\Big|_{\tiny\begin{matrix}p^2=-\mu_R^2 \\ \!\!\!\!p_z=p^R_z\end{matrix}},\\
&Z_{\slashed{p}}(z, p^R_z, a^{-1}, \mu_R)\nonumber\\
&\equiv \Big[\widetilde{F}_t(p,z)+\widetilde{F}_z(p,z)+\widetilde{F}_p(p,z)\Big]\bigg|_{\tiny\begin{matrix}p^2=-\mu_R^2 \\ \!\!\!\!p_z=p^R_z\end{matrix}}.
\end{align}

The bare nucleon matrix element from a Lattice calculation in coordinate space
\begin{align}
\widetilde{h}(z,P_z,a^{-1}) = {1\over 2P^0} \langle P |O_{\gamma^t}(z)| P \rangle \nn
\end{align}
is renormalized according to
\begin{align} \label{eq:rimomh}
&\widetilde{h}_R(z,P_z, p^R_z,\mu_R)\nn\\
&=\left.Z^{-1}(z,p^R_z,a^{-1},\mu_{\tiny R})\widetilde{h}(z,P_z,a^{-1})\right|_{a\to 0} .
\end{align}
Here $\widetilde{h}_R(z,P_z, p^R_z,\mu_R)$ is the continuum limit of the renormalized matrix element. Consequently, the quasi-PDF $\widetilde{q}_R(x,P_z,p^R_z,\mu_R)$ in  RI/MOM scheme is obtained through the Fourier transformation of $\widetilde{h}_R(z,P_z, p^R_z,\mu_{\tiny R})$:
\begin{align}
\widetilde{q}_R(x,P_z, p^R_z,\mu_R) = P_z\int {dz\over 2\pi}\ e^{ix P_zz}\widetilde{h}_R(z,P_z, p^R_z,\mu_R).\nn\\
\end{align}
In RI/MOM,
$\widetilde{h}_R(z,P_z, p^R_z,\mu_R)$ and $\widetilde{q}_R(x,P_z, p^R_z,\mu_R)$ are independent of the UV regulator, and the one-step matching between the quasi-PDF and $\overline{\rm MS}$ PDF can be carried out in the continuum theory with dimensional regularization~\cite{Stewart:2017tvs}.

The quasi-PDFs will eventually be matched to the same $\overline{\text{MS}}$ PDF, and  the two projections with
$Z_{mp}$ and $Z_{\slashed p}$  should  generate the same result. However the matching coefficient
can only be calculated at a fixed loop order, hence remanent  dependence on the projection operator is inevitable.

Using the same logic one reaches the conclusion that the quasi-PDF's dependence on the RI/MOM scales $\mu_R$ and $p_z^R$  should also be fully cancelled by the matching coefficient. Any fixed-order matching calculation will inevitably lead to a residual  $\mu_R$, $p_z^R$, and $P_z$  dependence of the final result for the PDF. These dependencies should  be carefully studied and included in the  systematic uncertainties.

\subsection{One-loop matching for quasi-PDF and PDF}\label{sec:one_loop_matching}

To obtain the matching coefficient  between the quasi-PDF $\widetilde{q}_R(x,P_z, p^R_z,\mu_R)$ and lightcone PDF $q(x,\mu)$, one  can calculate the  off-shell quark matrix elements in perturbation theory. In the following,  the calculation  will be conducted  in Landau gauge for both   minimal and $\slashed p$ projections. See Appendix~\ref{app:one-loop} for the results in a general covariant gauge with a general Lorentz structure.

The lowest order quark quasi-PDF is
\begin{equation}
\widetilde{q}^{(0)}(x)=\delta(1-x)\,.
\end{equation}
\begin{widetext}
At one-loop order, it is
\begin{align}\label{eq:one_loop_quasiPDF}
\widetilde{q}^{(1)}(x,p,\rho)=\mbox{Tr}\left[\left(\left[\widetilde{f}_t(x,\rho)\right]_+\gamma^t+\left[\widetilde{f}_z(x,\rho)\right]_+\frac{p_t}{p_z}\gamma^z+\left[\widetilde{f}_p(x,\rho)\right]_+\frac{p_t\slashed{p}}{p^2}\right){\cal P}\right]\,.
\end{align}
The $\widetilde{f}_i$s are
\begin{align}
\widetilde{f}_t(x,\rho)=\frac{\alpha_s C_F}{2\pi}\left\{
\begin{array}{lc}
\frac{8x^2(1-x)-x\rho(13-10x)+3\rho^2}{2(1-x)(1-\rho)(4x-4x^2-\rho)}+\frac{4x(2-x)-x\rho-3\rho}{4(1-x)(1-\rho)^{3/2}}\ln\frac{2x-1+\sqrt{1-\rho}}{2x-1-\sqrt{1-\rho}} & x>1\\
\frac{x(-7+4x)+3\rho}{2(1-x)(1-\rho)}+\frac{4x(2-x)-\rho(3+x)}{4(1-x)(1-\rho)^{3/2}}\ln\frac{1+\sqrt{1-\rho}}{1-\sqrt{1-\rho}} & 0<x<1\\
-\frac{8x^2(1-x)-x\rho(13-10x)+3\rho^2}{2(1-x)(1-\rho)(4x-4x^2-\rho)}-\frac{4x(2-x)-x\rho-3\rho}{4(1-x)(1-\rho)^{3/2}}\ln\frac{2x-1+\sqrt{1-\rho}}{2x-1-\sqrt{1-\rho}} & x<0
\end{array}\right.,
\end{align}
\begin{align}
\widetilde{f}_z(x,\rho)=\frac{\alpha_s C_F}{2\pi}\left\{
\begin{array}{lc}
\begin{array}{l}
\frac{-32x^2(1-x)^2(2x-1)-4x\rho(8-43x+65x^2-38x^3+8x^4)+\rho^2(5-41x+42x^2-8x^3)+2\rho^3(2-x)}{2(1-x)(1-\rho)^2(4x-4x^2-\rho)^2}\\
\quad+\frac{4-8x+8x^2+\rho(3-13x+4x^2)+2\rho^2}{4(1-x)(1-\rho)^{5/2}}\ln\frac{2x-1+\sqrt{1-\rho}}{2x-1-\sqrt{1-\rho}}
\end{array} & x>1\\
\frac{-5+15x-12x^2-2\rho(2-3x)}{2(1-x)(1-\rho)^2}+\frac{4-8x+8x^2+\rho(3-13x+4x^2)+2\rho^2}{4(1-x)(1-\rho)^{5/2}}\ln\frac{1+\sqrt{1-\rho}}{1-\sqrt{1-\rho}} & 0<x<1\\
\begin{array}{l}
-\frac{-32x^2(1-x)^2(2x-1)-4x\rho(8-43x+65x^2-38x^3+8x^4)+\rho^2(5-41x+42x^2-8x^3)+2\rho^3(2-x)}{2(1-x)(1-\rho)^2(4x-4x^2-\rho)^2}\\
\quad-\frac{4-8x+8x^2+\rho(3-13x+4x^2)+2\rho^2}{4(1-x)(1-\rho)^{5/2}}\ln\frac{2x-1+\sqrt{1-\rho}}{2x-1-\sqrt{1-\rho}}
\end{array} & x<0
\end{array}\right.,
\end{align}
\begin{align}
\widetilde{f}_p(x,\rho)=\frac{\alpha_s C_F}{2\pi}\left\{
\begin{array}{lc}
\frac{16x\rho(1-x)^2(1-6x)-2\rho^2(1-22x+26x^2-4x^3)-\rho^3(7-6x)}{2(1-\rho)^2(4x-4x^2-\rho)^2}+\frac{-\rho(8-12x+\rho)}{4(1-\rho)^{5/2}}\ln\frac{2x-1+\sqrt{1-\rho}}{2x-1-\sqrt{1-\rho}} & x>1\\
\frac{2-4x+\rho(7-8x)}{2(1-\rho)^2}+\frac{-\rho(8-12x+\rho)}{4(1-\rho)^{5/2}}\ln\frac{1+\sqrt{1-\rho}}{1-\sqrt{1-\rho}} & 0<x<1\\
-\frac{16x\rho(1-x)^2(1-6x)-2\rho^2(1-22x+26x^2-4x^3)-\rho^3(7-6x)}{2(1-\rho)^2(4x-4x^2-\rho)^2}-\frac{-\rho(8-12x+\rho)}{4(1-\rho)^{5/2}}\ln\frac{2x-1+\sqrt{1-\rho}}{2x-1-\sqrt{1-\rho}} & x<0
\end{array}\right.,
\end{align}
\end{widetext}
and
\begin{align}
\rho=\frac{-p^2-i\epsilon}{p_z^2}
\end{align}
with $i\epsilon$ giving the prescription to analytically extrapolate  $\rho$ from $\rho<1$ (Minkowski) to $\rho>1$ (Euclidean).  Notice that the vector current conservation guarantees  that  vertex corrections and  wave function contributions can  be combined  into    generalized plus functions~\cite{Stewart:2017tvs}. These functions are  defined with two arbitrary functions $h(x)$ and $g(x)$: 
\begin{align}
\int dx\left[{h(x)}\right]_+ g(x)&=\int dx\;h(x)\left[g(x)-g(1)\right]\,.
\end{align}

For the lightcone PDF with the same off-shell IR regulation in   Landau gauge, the tree level contribution is
\begin{align}
q^{(0)}(x)=\delta(1-x)\,,
\end{align}
and the   one-loop correction in the $\overline{\text{MS}}$ scheme is
%\begin{widetext}
\begin{align}\label{eq:one_loop_lightconePDF}
&q^{(1)}(x,p,\mu)\nn\\
=&\mbox{Tr}\left[\left(\left[f_+\left(x,\frac{\mu^2}{p^2}\right)\right]_+\gamma^++\left[f_p(x)\frac{p^+\slashed{p}}{p^2}\right]_+\right) {\cal P}\right]\,.
\end{align}
Here
\begin{align}
f_+\left(x,\frac{\mu^2}{p^2}\right)=&\frac{\alpha_s C_F}{2\pi}\theta(x)\theta(1-x)\left[
\frac{-5+10x-6x^2}{2(1-x)}\right.\nn\\
&\left.+\frac{1+x^2}{1-x}\ln\frac{\mu^2}{-x(1-x)p^2}\right]\,, \\
f_p(x)=&\frac{\alpha_s C_F}{2\pi}(1-2x)\theta(x)\theta(1-x)\,.
\end{align}

To match the quasi-PDF to lightcone PDF, one needs to take the on shell limit ($p^2\to 0$ or $\rho\to0$) and the large momentum limit ($p_t\to p_z$) for the bare quasi-PDF
\begin{equation}
\widetilde{q}^{(1)}_B(x,\rho)=\widetilde{q}^{(1)}(x,(p_t\to p_z,\vec{p}_\perp,p_z),\rho\to 0).
\end{equation}
One can observe that both terms proportional to $\gamma^t$ and $\gamma^z$ in Eq. (\ref{eq:one_loop_quasiPDF}) approach lightcone operators in the large momentum limit  and the combination of them captures  the correct collinear behavior. Therefore the bare quasi-PDF in minimal projection is defined  to pick up the coefficient of $\gamma^t$ and $\gamma^z$ in Eq. (\ref{eq:one_loop_quasiPDF}):
\begin{align}
\widetilde{q}_B^{(1)}(x,\rho)\Big|_{mp}=\left[\widetilde{f}_t(x,\rho)+\widetilde{f}_z(x,\rho)\right]_+\bigg|_{\rho\to 0}\,.
\end{align}

For the lightcone PDF,  the coefficient of $\gamma^+$ in Eq. (\ref{eq:one_loop_lightconePDF}) is used for minimal projection:
\begin{align}
q^{(1)}(x,p,\mu)\Big|_{mp}= f_+\left(x,\frac{\mu^2}{p^2}\right)_+\,.
\end{align}
The bare matching coefficient is then derived as
\begin{align} \label{eq:mpbarematching}
f_{1,mp}\left(x,\frac{p_z}{\mu}\right)_+= \widetilde{q}^{(1)}_B(x,\rho)\Big|_{mp}-q^{(1)}(x,p,\mu)\bigg|_{mp},
\end{align}
where
\begin{align}
&f_{1,mp}\left(x,\frac{p_z}{\mu}\right)=\frac{\alpha_s C_F}{2\pi} \nn\\
&\times \left\{
\begin{array}{lc}
\displaystyle \frac{1+x^2}{1-x}\ln\frac{x}{x-1}+1 & x>1\\
\displaystyle \frac{1+x^2}{1-x}\ln\frac{4x(1-x)p_z^2}{\mu^2}-\frac{x(1+x)}{1-x} & 0<x<1\\
\displaystyle -\frac{1+x^2}{1-x}\ln\frac{x}{x-1}-1 & x<0
\end{array} \right. .
\end{align}

In  RI/MOM, the  quasi-PDF is renormalized with  an additional counterterm. We find that in the $|x|\to\infty$  limit , only $\widetilde{f}_t(x,\rho)$ behaves as $1/|x|$. When integrating over $x$, this term  recovers   UV divergence in the local limit $z=0$. Therefore, it is a natural choice to pick up the  $\gamma^t$ term in Eq. (\ref{eq:one_loop_quasiPDF}) as a counterterm: 
\begin{align}
\widetilde{q}_{CT}^{(1)}\left(x,r,\frac{p_z}{p_z^R}\right)\bigg|_{mp}&=\left[\left|\frac{p_z}{p_z^R}\right|f_{2,mp}\left(1+\frac{p_z}{p_z^R}(x-1),r\right)\right]_+\,.
\end{align}
Here $r=\mu_R^2/(p_z^R)^2$, and
\begin{widetext}
\begin{align}
f_{2,mp}(x,r)=\widetilde{f}_t(x,r)=\frac{\alpha_s C_F}{2\pi}\left\{
\begin{array}{lc}
\frac{-3r^2+13rx-8x^2-10rx^2+8x^3}{2(r-1)(x-1)(r-4x+4x^2)}+\frac{-3r+8x-rx-4x^2}{2(r-1)^{3/2}(x-1)}\tan^{-1}\frac{\sqrt{r-1}}{2x-1} & x>1\\
\frac{-3r+7x-4x^2}{2(r-1)(1-x)}+\frac{3r-8x+rx+4x^2}{2(r-1)^{3/2}(1-x)}\tan^{-1}\sqrt{r-1} & 0<x<1\\
-\frac{-3r^2+13rx-8x^2-10rx^2+8x^3}{2(r-1)(x-1)(r-4x+4x^2)}-\frac{-3r+8x-rx-4x^2}{2(r-1)^{3/2}(x-1)}\tan^{-1}\frac{\sqrt{r-1}}{2x-1} & x<0
\end{array} \right. .
\end{align}
Finally, the matching coefficient $C$ in  the factorization formula given in Eq.\eqref{eq:fact}  is derived  as
\begin{align}\label{eq:matching_coeff}
C\left(x,r,\frac{p_z}{\mu},\frac{p_z}{p_z^R}\right)&=\delta(1-x)+\left[\widetilde{q}^{(1)}_B(x,\rho)-q^{(1)}(x,p,\mu)-\widetilde{q}_{CT}^{(1)}\left(x,r,\frac{p_z}{p_z^R}\right)\right]\Bigg|_{mp}+{\cal O}(\alpha_s^2)\nonumber\\
&=\delta(1-x)+\left[f_{1,mp}\left(x,\frac{p_z}{\mu}\right)-\left|\frac{p_z}{p_z^R}\right|f_{2,mp}\left(1+\frac{p_z}{p_z^R}(x-1),r\right)\right]_+ +{\cal O}(\alpha_s^2)\,.
\end{align}
\end{widetext}
Here the coupling $\alpha_s(\mu)$ is in the standard $\overline{\text{MS}}$ scheme. Note that   the antiquark distribution is mapped  into the region $-1<y<0$ by setting $q(y)=-\bar{q}(-y)$.

For $\slashed{p}$ projection, one has the bare quasi-PDF
\begin{align}
\widetilde{q}_B^{(1)}(x,\rho)\Big|_{\slashed{p}}=\left[\widetilde{f}_t(x,\rho)+\widetilde{f}_z(x,\rho)+\widetilde{f}_p(x,\rho)\right]_+\,.
\end{align}
The  lightcone PDF with a similar projection is
\begin{align}
q^{(1)}(x,p,\mu)\Big|_{\slashed{p}}=\left[f_+\left(x,\frac{\mu^2}{p^2}\right)+f_p(x)\right]_+\,.
\end{align}
Under this projection, the matching coefficient for the bare quasi-PDF coincides with Eq.~(\ref{eq:mpbarematching}):
\begin{align}
f_{1,\slashed p}\left(x,\frac{p_z}{\mu}\right)_+&= \widetilde{q}^{(1)}_B(x,\rho)\Big|_{\slashed p}-q^{(1)}(x,p,\mu)\bigg|_{\slashed p} \nn\\
&=f_{1,mp}\left(x,\frac{p_z}{\mu}\right)_+\,.
\end{align}
The counter-term  can  be obtained by  calculating   with ${\cal P}=\slashed{p}/(4p^t)$:
\begin{widetext}
\begin{align}
f_{2,\slashed{p}}(x,r)=\frac{\alpha_s C_F}{2\pi}\left\{
\begin{array}{lc}
\frac{3-3r-2x}{2(r-1)(x-1)}+\frac{4rx-8x^2+8x^3}{(r-4x+4x^2)^2}+\frac{2-2r-rx+2x^2}{(r-1)^{3/2}(x-1)}\tan^{-1}\frac{\sqrt{r-1}}{2x-1} & x>1\\
\frac{3-3r-2x+4x^2}{2(r-1)(1-x)}+\frac{-2+2r+rx-2x^2}{(r-1)^{3/2}(1-x)}\tan^{-1}\sqrt{r-1} & 0<x<1\\
-\frac{3-3r-2x}{2(r-1)(x-1)}-\frac{4rx-8x^2+8x^3}{(r-4x+4x^2)^2}-\frac{2-2r-rx+2x^2}{(r-1)^{3/2}(x-1)}\tan^{-1}\frac{\sqrt{r-1}}{2x-1} & x<0
\end{array} \right.\,.
\end{align}
\end{widetext}
The corresponding RI/MOM matching coefficient is obtained by replacing ``$mp$" with ``$\slashed p$" in Eq.~(\ref{eq:matching_coeff}),  {and the difference between $f_{2,\slashed{p}}$ and $f_{2,mp}$ vanishes in the $p_z^R$=0 limit.} The matching coefficient with $\Gamma=\gamma^z$ is also given in Appendix \ref{app:one-loop_gamma_z}.

\section{Lattice calculation of PDF}\label{sec:numerical}

\subsection{Lattice Matrix Elements}

In this subsection,  we give  the results of a Lattice-QCD calculation using clover valence fermions on  {the CLS $32^3\times96$ 2+1 flavor clover fermion ensemble H102 with Lattice spacing $a=0.086$~fm, pion mass $M_\pi=$ 356~MeV and box size $L\approx 2.7$~fm ($M_\pi L\approx 4.9$)~\cite{Bruno:2014jqa}. We use  $\kappa_l=0.136865$ and $C_{SW}=1.98625$ for the valence clover fermion. We apply APE smearing~\cite{Hasenfratz:2001hp} with size=2.5$a$ twice in the source/sink smearing and also in the quasi-PDF operator $O_\Gamma$, but not in the fermion propagators.}

\begin{table}[htbp]
\begin{center}
\caption{\label{table:p_modes}
Momentum modes   used in the NPR analysis. The four digits (in units of $2\pi/L$) in   brackets correspond to   three spatial momentum components and the energy  component. The $z$ direction can be selected by  setting   the Wilson link along any of the three spatial directions.  Thus these momentum modes approximately cover  three choices  of $\mu_R=\sqrt{-(p^R)^2}$ and several sets of $p^R_z= 2\pi i/L(i=0,1,2,...)$.  }
\begin{tabular}{c|ccc}
\hline 
$a^2\mu_R^2$ & Momentum modes $p^R$ \\
$ [1.109,1.118]$ &(5,2,0,0)   (4,3,1,5)   \\
$ [1.957,1.966]$ &(5,5,0,3)  (6,3,2,4)   (5,4,1,9) \\
$ [2.814,2.832]$ &(6,5,1,10)   (7,4,2,6)   (6,3,0,16)\\
\hline
\end{tabular}
\end{center}
\end{table}

First of all, we  will explore   the nonperturbative renormalization   in  RI/MOM scheme.  Following  Ref.~\cite{Chen:2017mzz} , we use Landau gauge fixed wall sources (while limiting the source in the  time slice range $t\in[32, 64]$ to avoid the boundary effect from   open boundary condition at $t$=0), and generate the propagators with the momentum modes listed in Table.~\ref{table:p_modes}.  The four digits in   brackets correspond to   three spatial component and the energy  component  in units of $2\pi/L$. The $z$ direction can be selected by  setting   the Wilson link along any of the three spatial directions.   Thus these results approximately  cover three values of  {$\mu_R=\sqrt{-(p^R)^2}$ (2.4, 3.2 and 3.9~GeV, corresponding to $a^2\mu_R^2$=1.1, 2.0 and 2.8), and     $p^R_z=\{0,1,2,...,\}*2\pi /L$} up to the upper limit $p^R_z<\mu_R$. Note that in deriving these momentum modes we have required  the spatial components of a given momentum mode different with  each other, and adjusted  $p_t^R$ to ensure  $\mu_R$ invariant (within 2\%). These  choices allow us to explore  the dependence on each component of $p$, but one should be cautious that  the results may suffer from sizable  discretization errors since the normal constraint $\frac{\sum_{\mu}a^4p_{\mu}^4}{(\sum_{\mu}a^2p_{\mu}^2)^2}<0.3$ is not respected. These  discretization errors will be investigated in the future.

\begin{figure}[htbp]
\includegraphics[width=.48\textwidth]{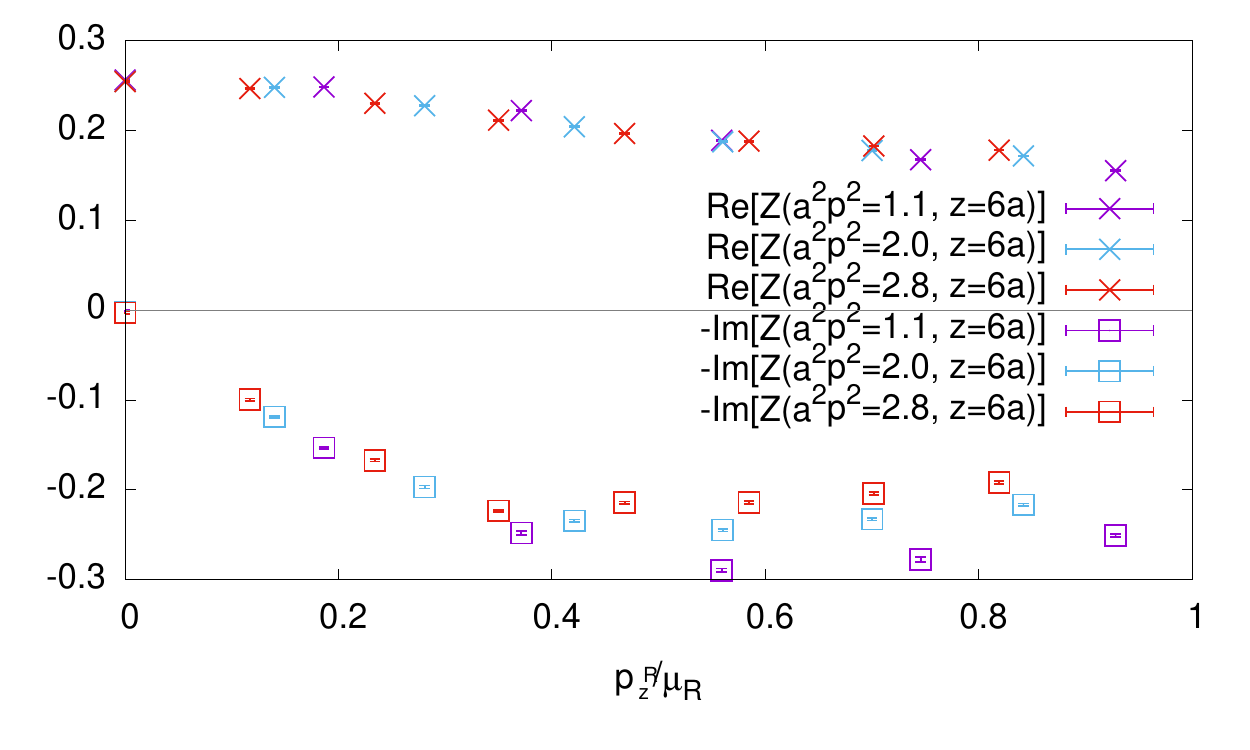}
\includegraphics[width=.48\textwidth]{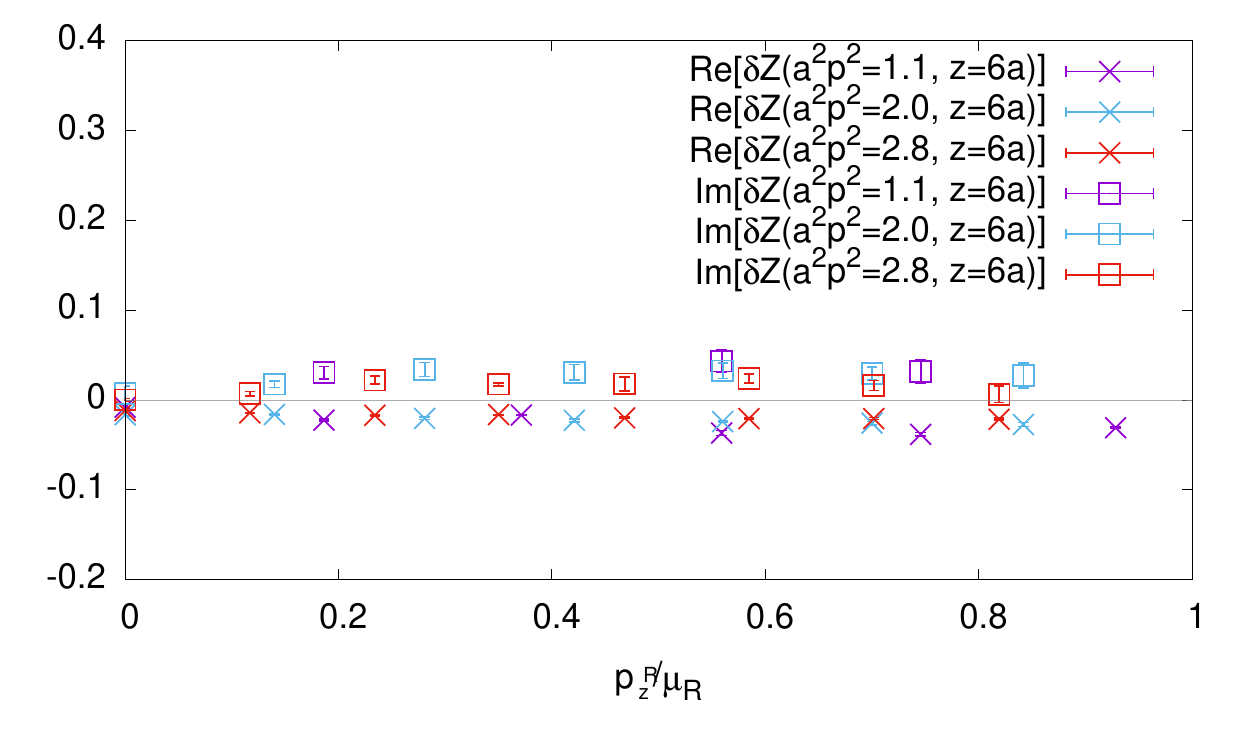}
\caption{The NPR  $Z=Z_{mp}$ (top) and $\delta Z=Z_{\slashed{p}}-Z_{mp}$ (bottom) at $z=6a$ ($\approx 0.5$~fm) as a function of $p_z^R/\mu_R=1/\sqrt{r}$, with various $\mu_R$.   At $p_z^R=0$, $Z$ is real and $\delta Z/Z$ is less than 5\%. } \label{fig:pz_dependence}
\end{figure}

As shown in Eq.~(\ref{eq:matching_coeff}), the one-loop matching formula   primarily depends on the combination $r=\mu^2_R/(p_z^R)^2$ but  is independent of $p^R_t$. In Fig.~\ref{fig:pz_dependence}, the  $Z_{mp}$ and $Z_{\slashed{p}}-Z_{mp}$ at fixed $z\sim 0.5$ fm are plotted as a function of $p^R_z/\mu_R=1/\sqrt{r}$. From this figure, one can see   that the NPR factors, both real   and imaginary parts,  only show the dependence on  $r$ regardless of the values of $p_z^R$ or $\mu_R$, with  $p_z^R/\mu_R<0.4$.

\begin{figure}[tbp]
\includegraphics[width=.48\textwidth]{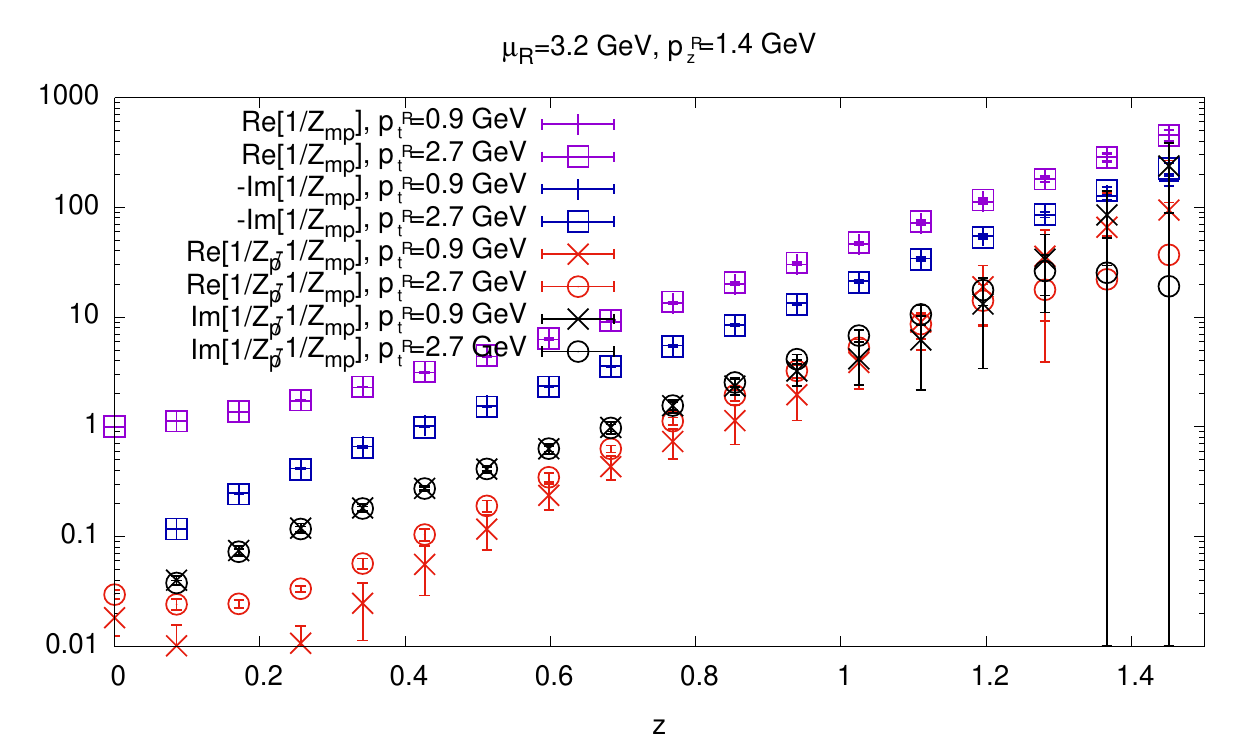}
\caption{The  inverse of the minimum projection renormalization factor $1/Z_{mp}$ and the difference $1/Z_{\slashed{p}}-1/Z_{mp}$, as a function of $z$ with the same $p^R_z$ and $\mu_R$, but different $p^R_t$.  The crosses correspond to $p_t^R=0.9$ GeV and the open boxes (circles) are for $p_t^R=2.7$ GeV. Most results show mild dependence on $p^R_t$.} \label{fig:pt_dependence}
\end{figure}

In Fig.~\ref{fig:pt_dependence},  we show $1/Z_{mp}$ and $1/Z_{\slashed{p}}-1/Z_{mp}$ as functions of the Wilson link length $z$, with the same $\mu_R$=3.2 GeV and $p_z^R=1.4$ GeV and two different values of $p_t^R$=0.9 GeV and 2.7 GeV.  As shown in this figure, the  $1/Z_{mp}$ and $1/Z_{\slashed{p}}-1/Z_{mp}$ with the two different $p_t^R$'s are close  to each other for all $z$ (the curves with the same color).    This  is consistent with the 1-loop matching formula.   At $z<0.3$ fm, the real part of $1/Z_{\slashed{p}}-1/Z_{mp}$  with the two different $p_t^R$'s can be slightly nonzero, but it is  still  smaller than $1/Z_{\slashed{p}}$ by  two orders of magnitude.

In the following  we will take $p_z^R$ to be zero and estimate the systematic uncertainty from the $p_z^R$ dependence by varying the $p_z^R$.

In the calculation of nucleon matrix element, we use Gaussian momentum smearing~\cite{Bali:2016lva} for the quark field
\begin{align}\label{eq:moms}
\psi(x)\to &S_\text{mom}\psi(x) = \frac{1}{1+6\alpha}\nonumber\\
&\left[\psi(x) + \alpha \sum_j U^{APE}_j(x)e^{ik\hat{e}_j}\psi(x+\hat{e}_j)\right]\, ,
\end{align}
where $k$ is the desired momentum,  {$U^{APE}_j(x)$ are the APE smeared} gauge links in the $j$ direction, and $\alpha$ is a tunable parameter as in traditional Gaussian smearing.

Such a momentum source is designed to increase the overlap with nucleons of the desired boost momentum and we are able to reach
higher-boosted momentum for the nucleon states than
in the  previous work~\cite{Chen:2017mzz}. Although in the  exploratory
study, we varied the  Gaussian smearing radius to better
overlap with the largest momentum used in the calculation, the field smearing  is still centered around
zero momentum in momentum space. When we switch
to the momentum smearing, the  smearing center will be shifted to momentum $O(k)$, which will immediately allow us to reach higher boost momenta with better signal-to-noise ratios in the matrix elements. 
In this work, we use  two  values of nucleon boost momenta, $P_z=n \frac{2\pi}{L}$, with $n \in \{4,5\}$, which corresponds to 1.8 and 2.3~GeV.

\begin{figure*}[htbp]
\includegraphics[width=.49\textwidth,height= 0.35\textwidth]{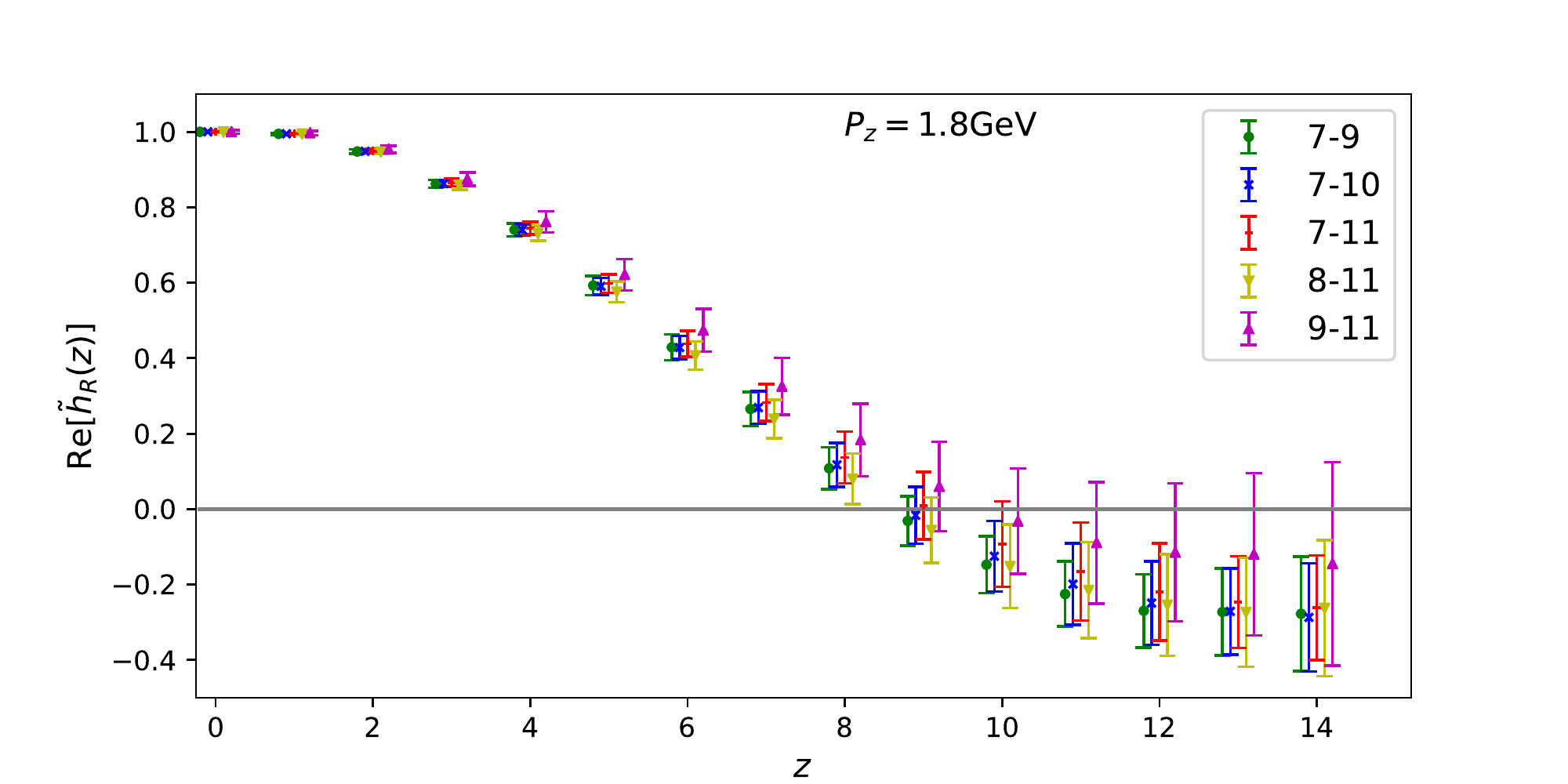}
\includegraphics[width=.49\textwidth,height= 0.35\textwidth]{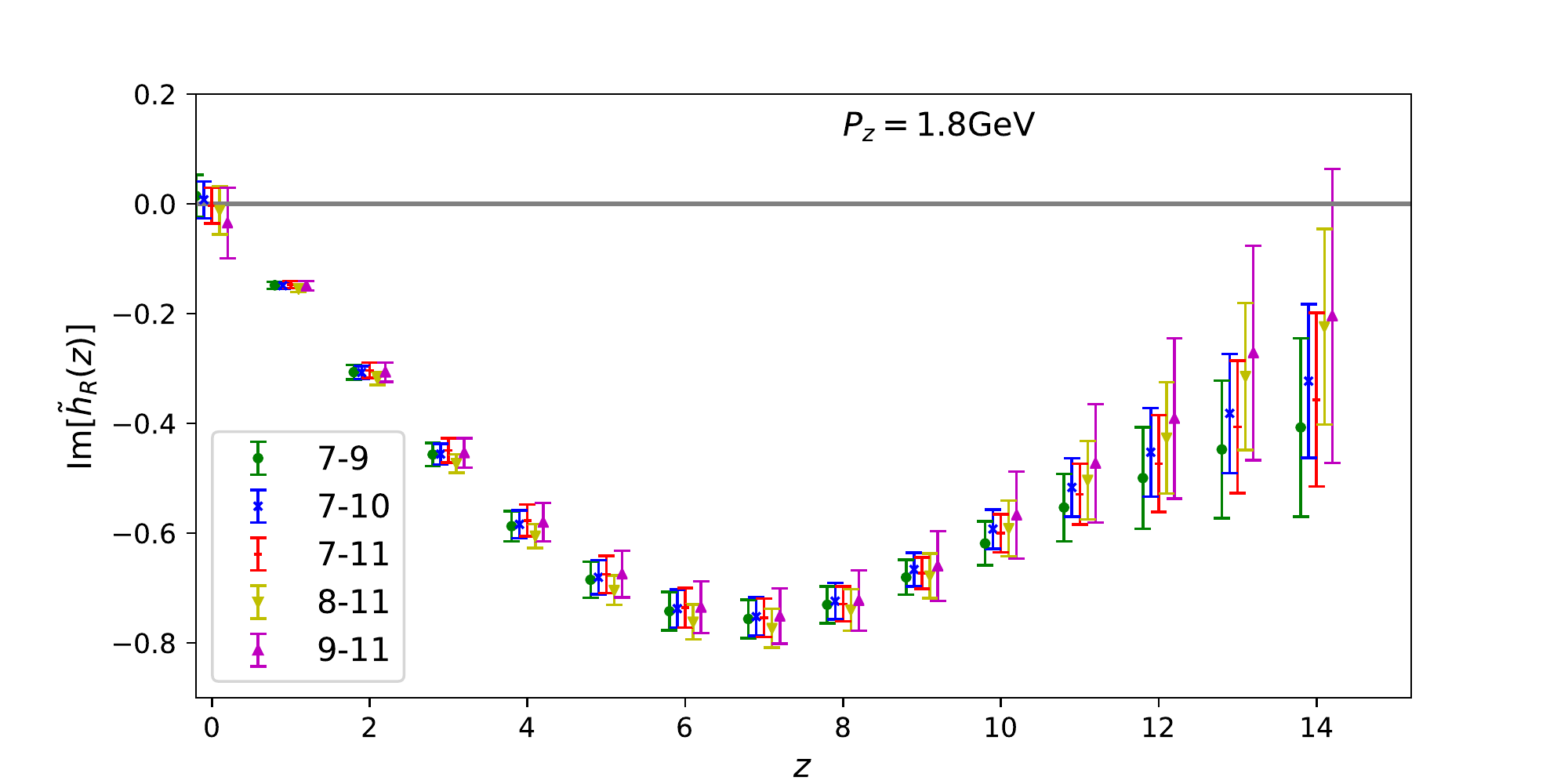}\\
\includegraphics[width=.49\textwidth,height= 0.35\textwidth]{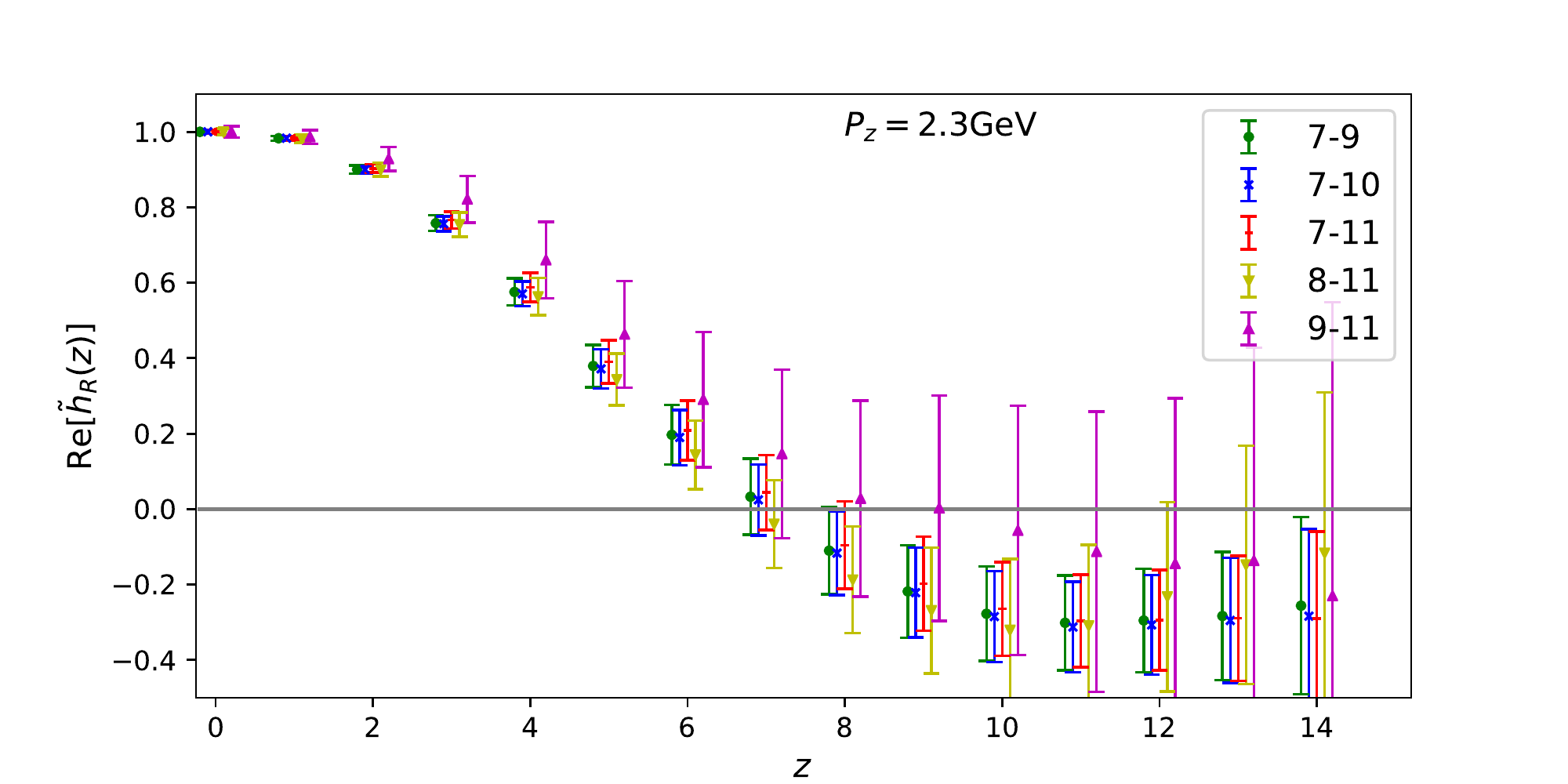}
\includegraphics[width=.49\textwidth,height= 0.35\textwidth]{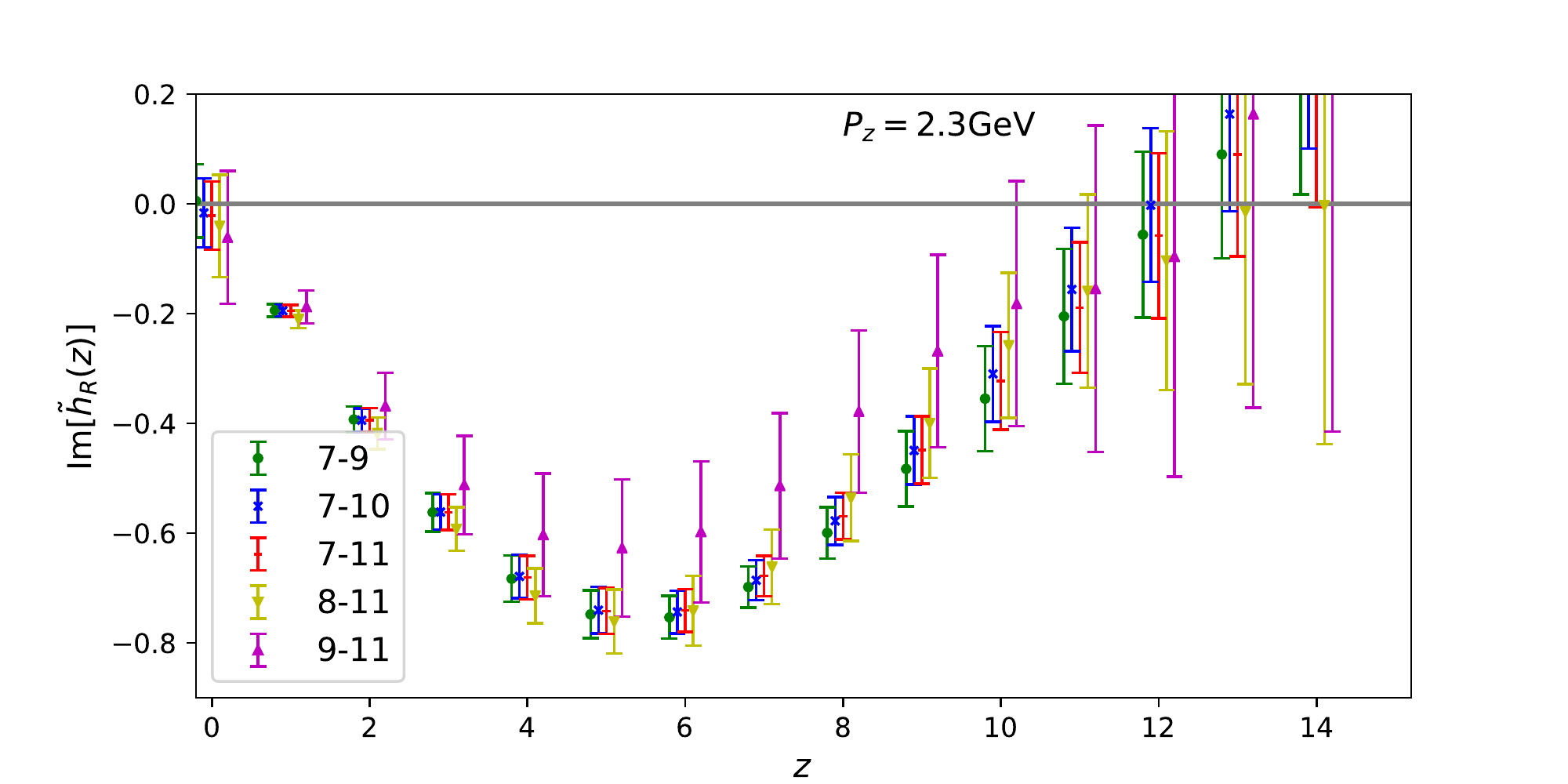}
\caption{The real (left) and imaginary (right) parts of the isovector nucleon matrix elements for unpolarized PDFs as functions of $z$ at different momenta, with $P_z=\frac{8\pi}{L}$=1.8 GeV (top) and $\frac{10\pi}{L}=$2.3~GeV (bottom) respectively. The RI/MOM renormalization factors with $\{\mu, p_z^R\}=\{3.2, 0\}$ GeV and   the normalization at $z=0$ are applied on the bare matrix elements to improve the visibility at large $z$.  At a given positive $z$ value, the data is slightly offset to show the ground-state matrix element from the fits using different ranges; from left to right they are: $t_\text{seq}\in$ [7,9], [7,10], [7,11], [8,11], and [9,11]. Different analyses are consistent within statistical errors while the fits with separation 7 and 8 have smaller  uncertainties compared  to   other cases.
}
\label{fig:bareME-tsep}
\end{figure*}

On the Lattice, we   calculate the time-independent and nonlocal  in $z$ direction  correlators of a nucleon with finite-$P_z$ boost
\begin{align}
\label{eq:qlat}
\widetilde{h}_\text{lat}(z,P_z,{\Gamma};a^{-1}) =
  \left\langle 0;\vec{P} \right|O_\Gamma(z)
  \left| 0;\vec{P} \right\rangle.
\end{align}
 {Here the state $|0; \vec{P}\rangle$ represents the ground (nucleon) state with momentum $\vec{P}=\{0,0,P_z\}$}.  $\Gamma=\gamma^t$ is used for the unpolarized parton distribution.

\begin{figure*}[htbp]
\includegraphics[width=.49\textwidth,height= 0.35\textwidth]{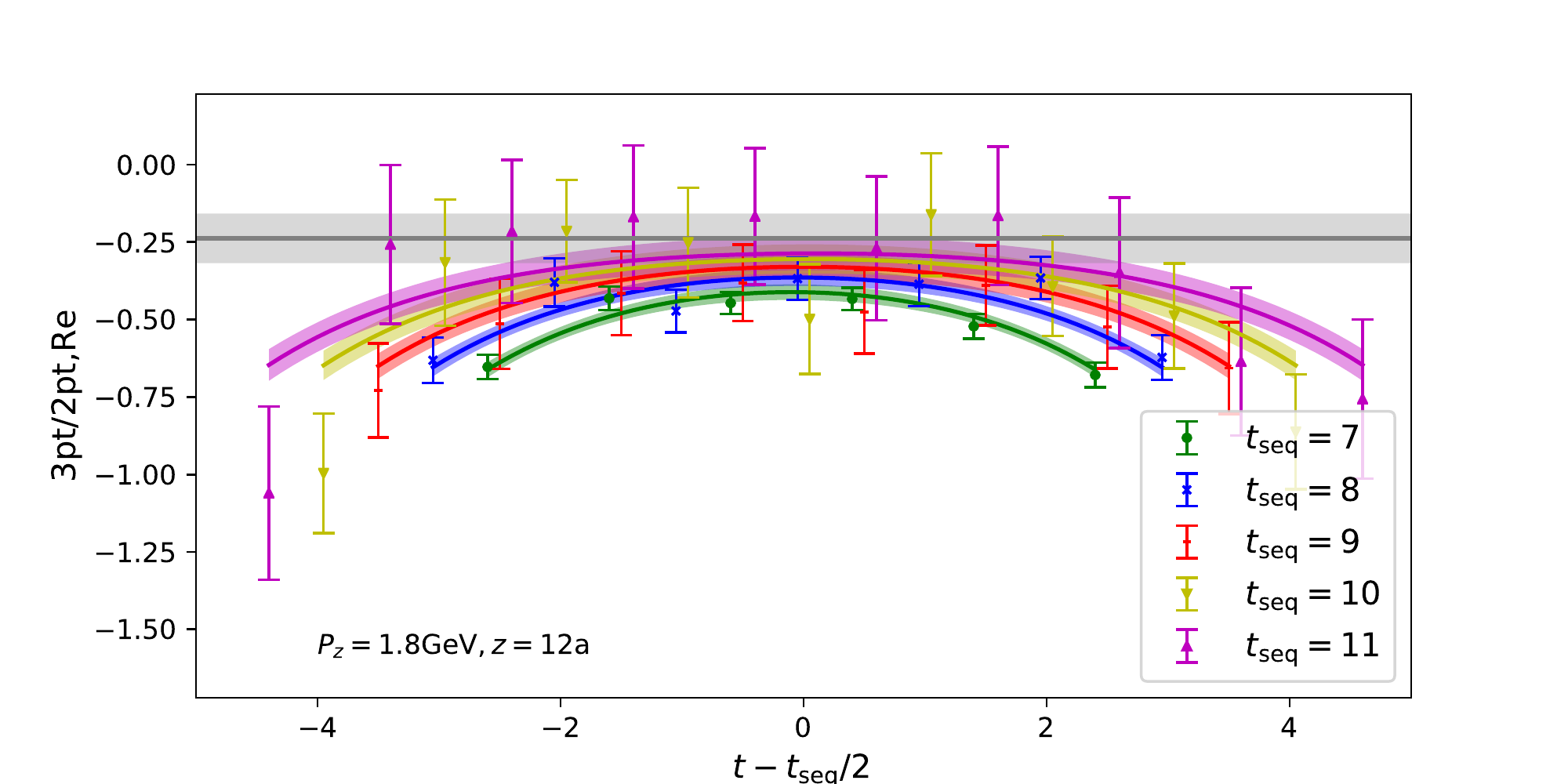}
\includegraphics[width=.49\textwidth,height= 0.35\textwidth]{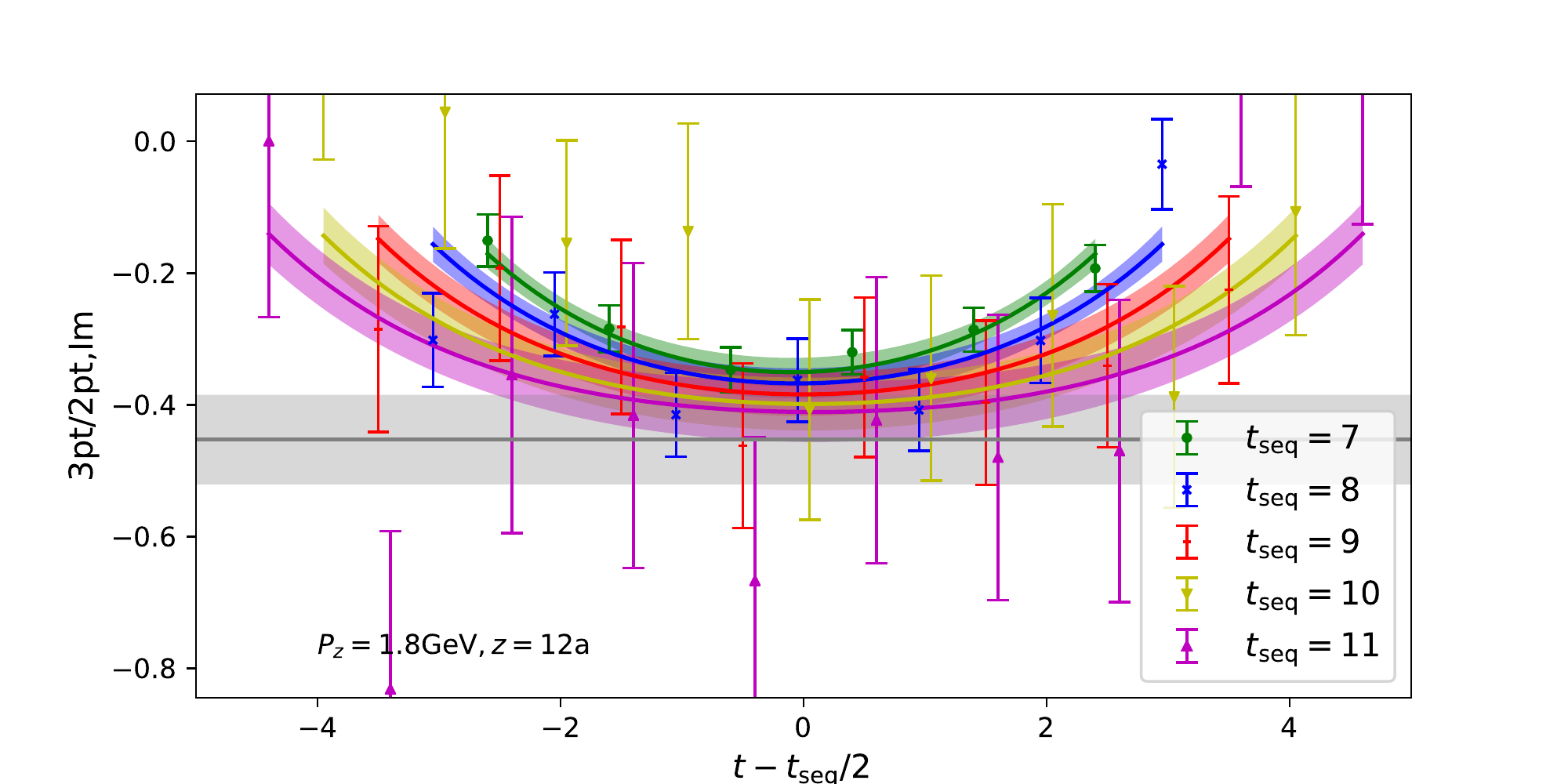}\\
\includegraphics[width=.49\textwidth,height= 0.35\textwidth]{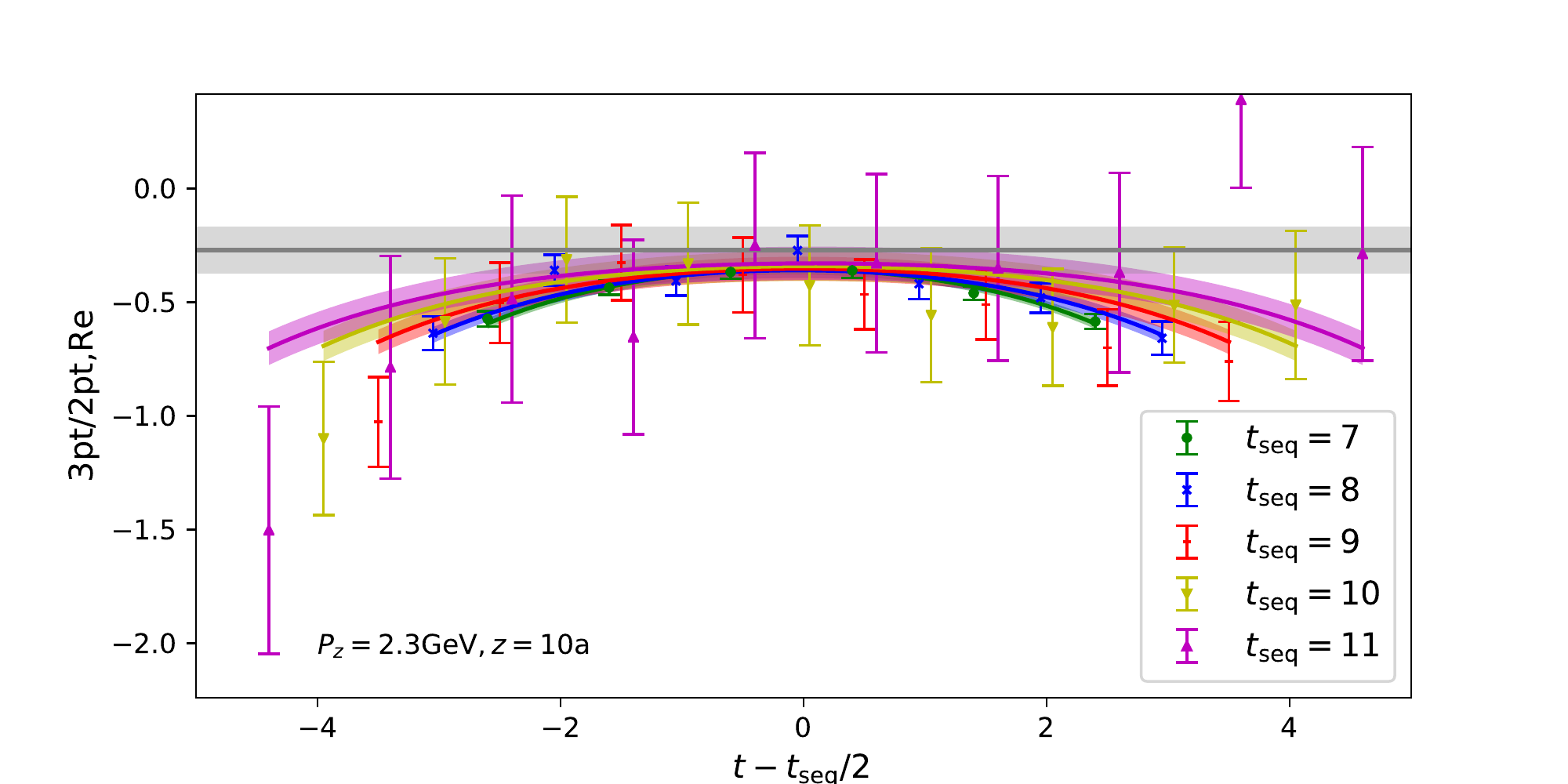}
\includegraphics[width=.49\textwidth,height= 0.35\textwidth]{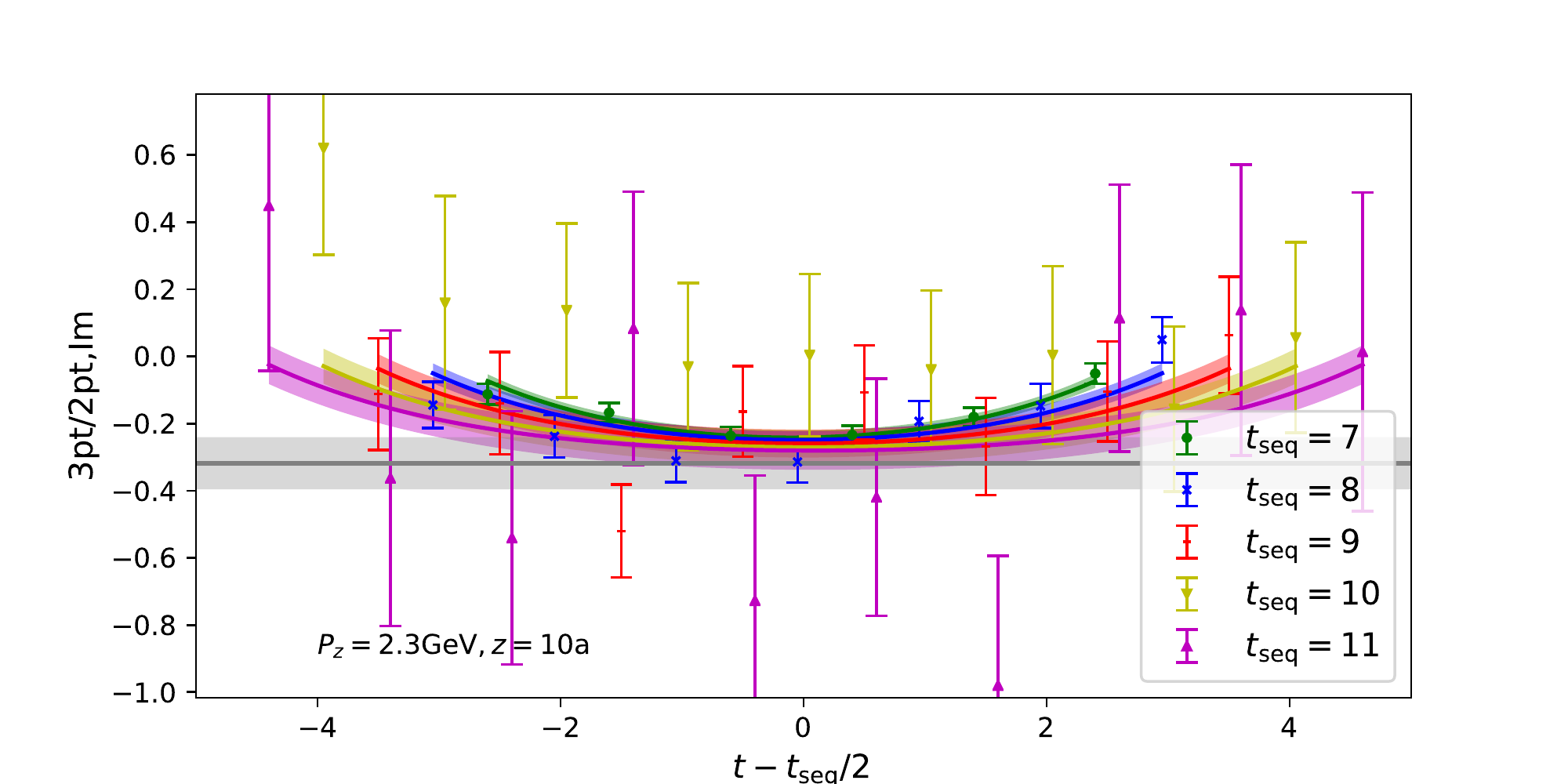}
\caption{The real (left) and imaginary (right) parts of the renormalized isovector nucleon matrix elements for unpolarized PDFs with ${P_z, z}={8\pi/L, 12}$ (top) and ${10\pi/L, 10}$ (bottom) which correspond to $zP_z\sim9.5$. The data points and the band predicted by the fit using $t_\text{seq}\in$ [7, 11] agree with each other well.}
\label{fig:bareME-tsep2}
\end{figure*}

\begin{figure}[tbp]
\includegraphics[width=0.48\textwidth,height= 0.3\textwidth]{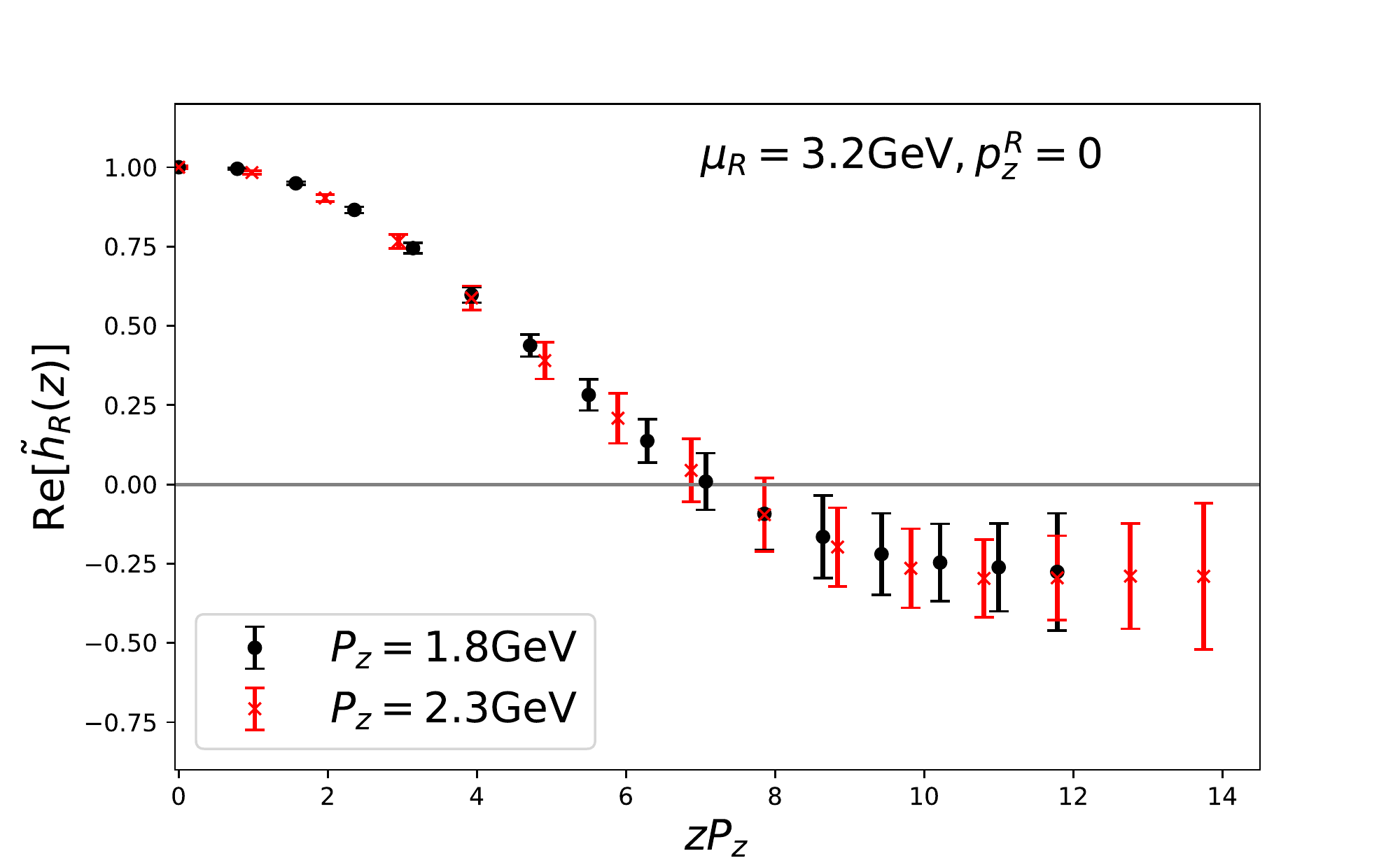}
\includegraphics[width=0.48\textwidth,height= 0.3\textwidth]{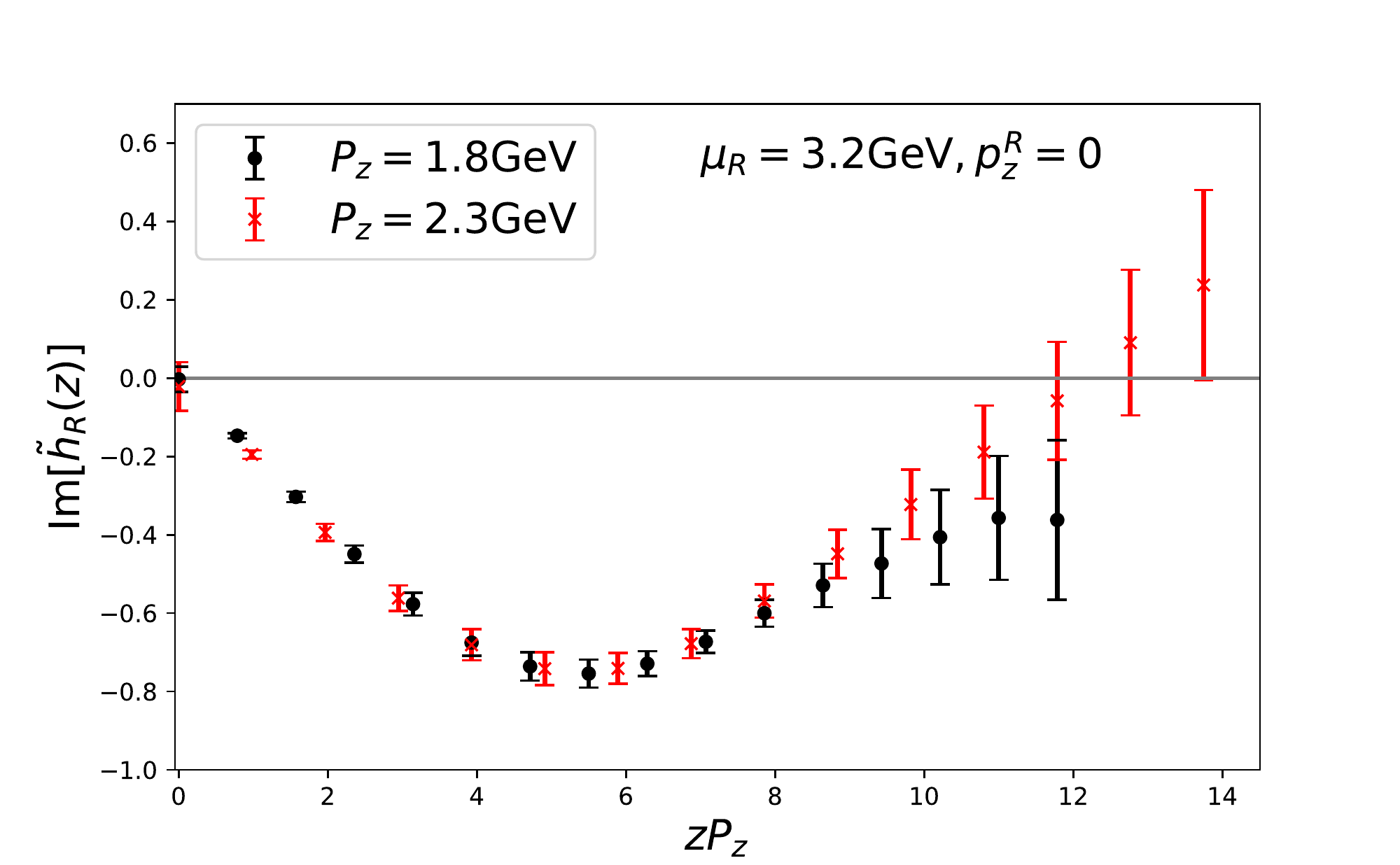}
\caption{The renormalized quasi-PDF matrix elements with $P_z=$1.8 and 2.3 GeV, using the minimal projection with $p_z^R=0$ and $\mu_R=3.2$ GeV, as   function of $zP_z$.} \label{fig:renormalized_h-tilde}
\end{figure}

As  the nucleon boost momentum increases, one anticipates that excited-state contributions  are more severe; therefore, a careful study of the excited-state contamination is necessary.
To do so, we calculate the nucleon matrix element $\widetilde{h}_{\rm lat}$ at five source-sink separations $t_\text{seq}\in\{7,8,9,10,11\}\times 0.086$~fm, with $\{4,4,8,8,16\}$~measurements on each of 2005 gauge configurations respectively in the $P_z=1.8$ GeV case, and  of 2,000 configurations in the $P_z=2.3$ GeV case. We use a multigrid algorithm~\cite{Babich:2010qb,Osborn:2010mb} with the Chroma software package~\cite{Edwards:2004sx} to speed up the inversion of the quark propagator.
Following  Ref.~\cite{Bhattacharya:2013ehc}, each three-point (3pt) correlator $C_\Gamma^{(3\text{pt})} (t, t_\text{seq})$ can be decomposed as (assuming the source is at $t= 0$)
\begin{align} \label{eq:fit}
C^\text{3pt}(t,t_\text{seq}; P_z, {\Gamma}) &=
   {\cal A}_0^2 \langle 0 | O_\Gamma | 0 \rangle  e^{-E_0t_\text{seq}} \\
   &+{\cal A}_1^2 \langle 1 | O_\Gamma | 1 \rangle  e^{-E_1t_\text{seq}} \nonumber\\
   &+{\cal A}_1{\cal A}_0 \langle 1 |O_\Gamma | 0 \rangle  e^{-E_1 (t_\text{seq}-t)} e^{-E_0 t} \nonumber\\
   &+{\cal A}_0{\cal A}_1 \langle 0 | O_\Gamma | 1 \rangle  e^{-E_0 (t_\text{seq}-t)} e^{-E_1 t} + \ldots \,,\nonumber
\end{align}
where  $|n\rangle$  with $n
> 0$ represents the excited states. The operator is inserted at time
$t$, and the nucleon state is annihilated at the sink time
$t_\text{seq}$( which is also the source-sink separation).
The spectrum weights ${\cal A}_{0,1}$ and energies $E_{0,1}$ in Eq.~(\ref{eq:fit}) can be obtained from the two-point (2pt) correlator: 
\begin{align} \label{eq:fit}
C^\text{2pt}(t_\text{seq}; P_z) &=
   {\cal A}_0^2 e^{-E_0t_\text{seq}} +{\cal A}_1^2 e^{-E_1t_\text{seq}} + \ldots\ .
\end{align}

Eventually we apply  the joint fit with the 3pt functions at several $t_\text{seq}$ and   2pt function using the following form~\cite{Bhattacharya:2013ehc}:
\begin{align}
&\frac{C^\text{3pt}(t,t_\text{seq})}{C^\text{2pt}(t_\text{seq})}= \nn\\
&=\frac{\tilde{h}_{\text{lat}}+C_2(e^{-\Delta Et}+e^{-\Delta E(t_\text{seq}-t)})+C_3e^{-\Delta E t_\text{seq}}}{1+C_1 e^{-\Delta Et_\text{seq}}},\nonumber\\
&C^\text{2pt}(t)=C_0e^{-E_0t}(1+C_1 e^{-\Delta Et}),
\end{align}
with $\Delta E= E_1-E_0$.  $C_{0,1,2,3}$ and $E_{0,1}$ are free parameters. We limit the range of $t$ as  $t\in[1,t_\text{seq}-1]$ for 3pt/2pt ratio and $t\in[7, 11]$ for the 2pt to make the $\chi^2/d.o.f.$ of the fit to be ${\cal O} (1)$.  Using the ratio of 3pt/2pt instead of the 3pt function itself can improve the stability of the fit, especially when $C^\text{2pt}(t)$ with $t<7$ is included in the fit.

In Fig.~\ref{fig:bareME-tsep}, we show the ground-state nucleon matrix elements $\tilde{h}_{lat}(z,P_z,\gamma_t)$ obtained from five fits: using the separations $t_\text{seq}\in$ [7, 9], [7, 10], [7,11], [8,11], and [9,11] (The data points correspond to the same $z$ but are shifted horizontally to enhance the visibility). The  data are further normalized by multiplying  the renormalization factor with $\{\mu_R, p_z^R\}=\{3.2, 0\}$ GeV and the real part normalized to 1 at $z=0$. From this figure, one can see that  there is no clear signal for excited-state contributions in any of these analyses.    If the data with    smallest two separations are dropped, uncertainties are getting much larger.     In the fit,  we keep the $C_3$ term to make a moderate estimate of the uncertainty even when this term is not statistically significant.

For  a comparison between data and the fit, we show our results at large $z$ like  $({P_z, z})=({8\pi/L, 12a})$ and $({10\pi/L, 10a})$ with  $t_\text{seq}\in$ [7,11] in Fig.~\ref{fig:bareME-tsep2}. In these  spatial separations,  the real part of  matrix element seems to be negative.   The ground-state contribution  obtained from the fit is shown as the black band. As one can see, most data can be well described in the fit and thereby we use the two-state fits and the interval  $t_\text{seq}\in$ [7, 11] to obtain the results   in the rest part of this paper.

The renormalized quasi-PDF matrix elements with two values of $P_z$ are plotted in Fig.~\ref{fig:renormalized_h-tilde}, as  function of $zP_z$ for  $p^R_z=0$ and $\mu_R=3.2$ GeV. The results with different $P_z$ are consistent with each other within statistical uncertainties.  This  indicates that power corrections due to   higher-twist effects might not be sizable.

\subsection{Systematic  {Uncertainties}}

\begin{figure}
\includegraphics[width=0.48\textwidth]{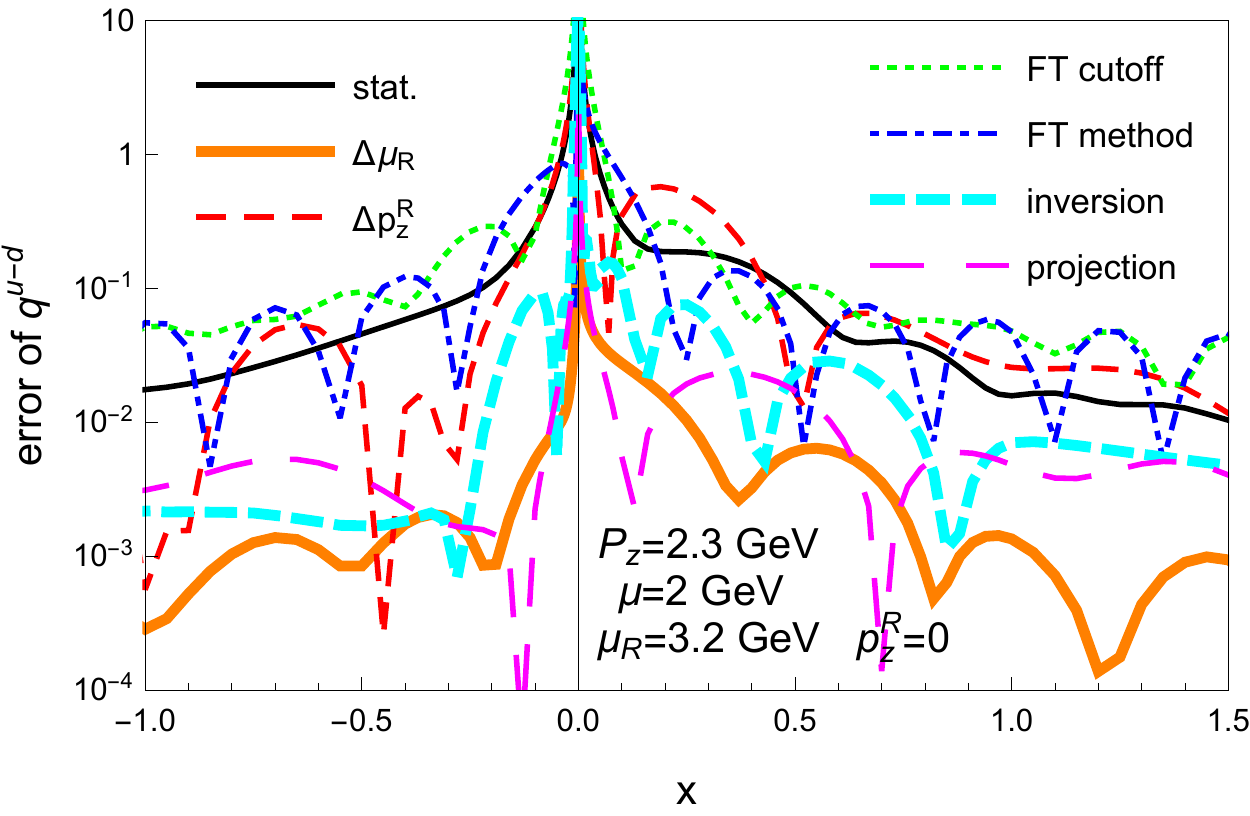}
\includegraphics[width=0.48\textwidth]{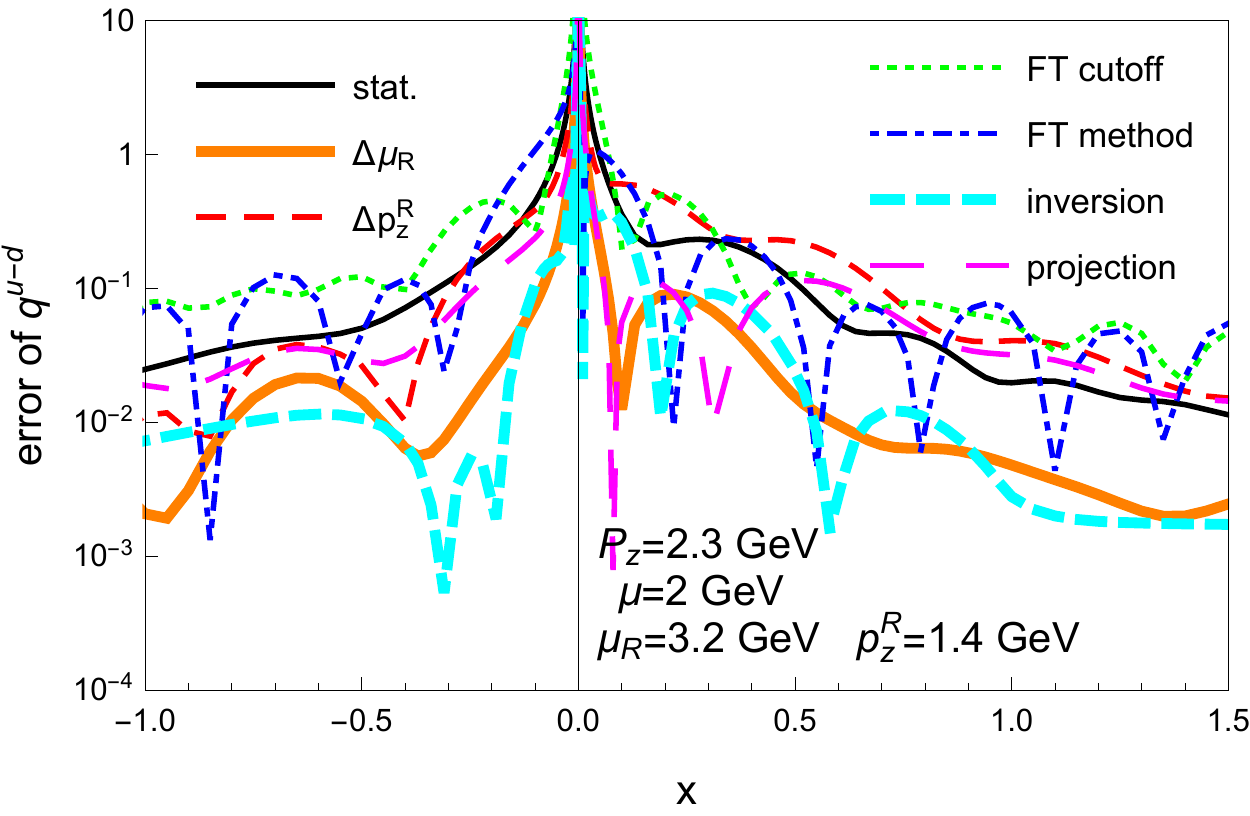}
\caption{\label{fig:error} Different contributions to the systematic errors. See the text for detailed information.  }
\end{figure}

In this subsection, we will consider four systematic uncertainties from: Fourier transformation (FT),  unphysical scales $p_z^R$ and $\mu_R$, projection used in the RI/MOM scheme, and inversion of the matching coefficient.

\begin{figure}[tbp]
\includegraphics[width=.48\textwidth]{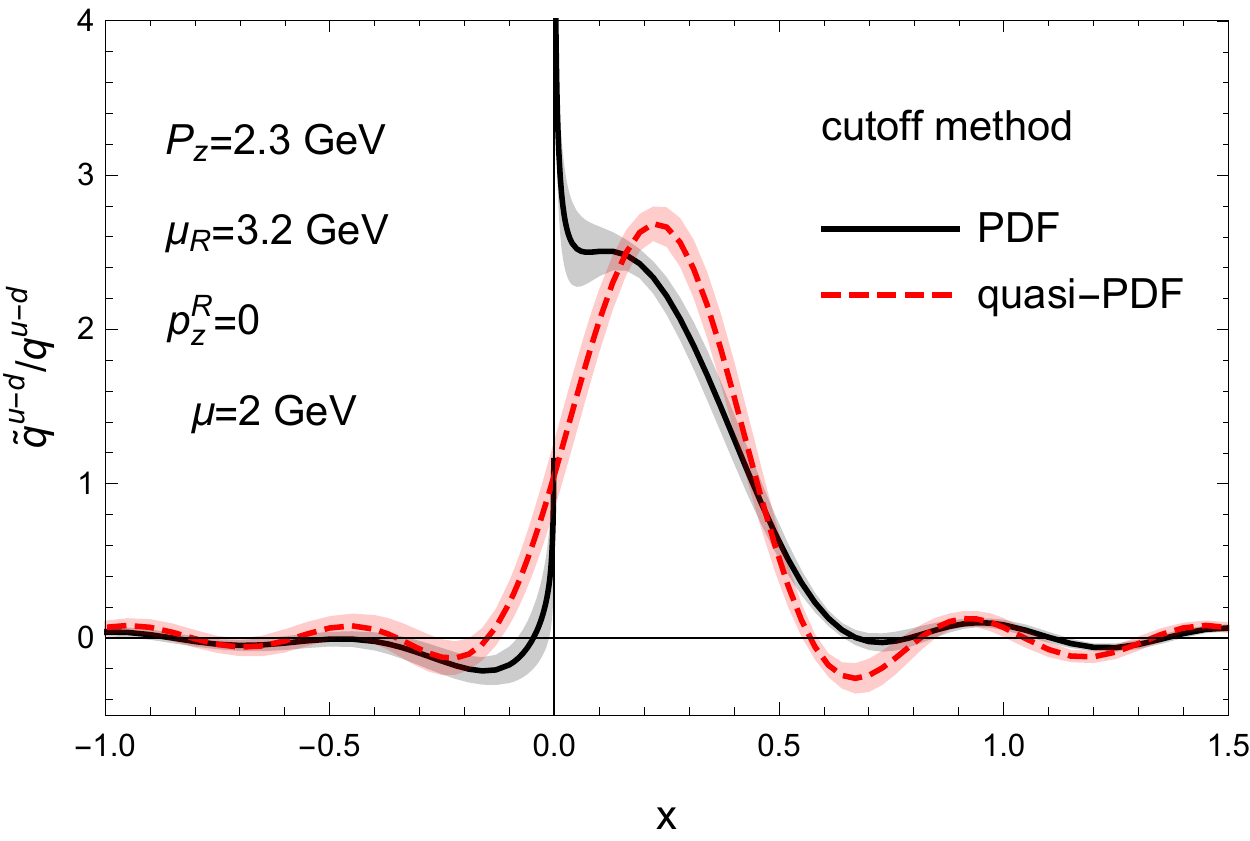}
\includegraphics[width=.48\textwidth]{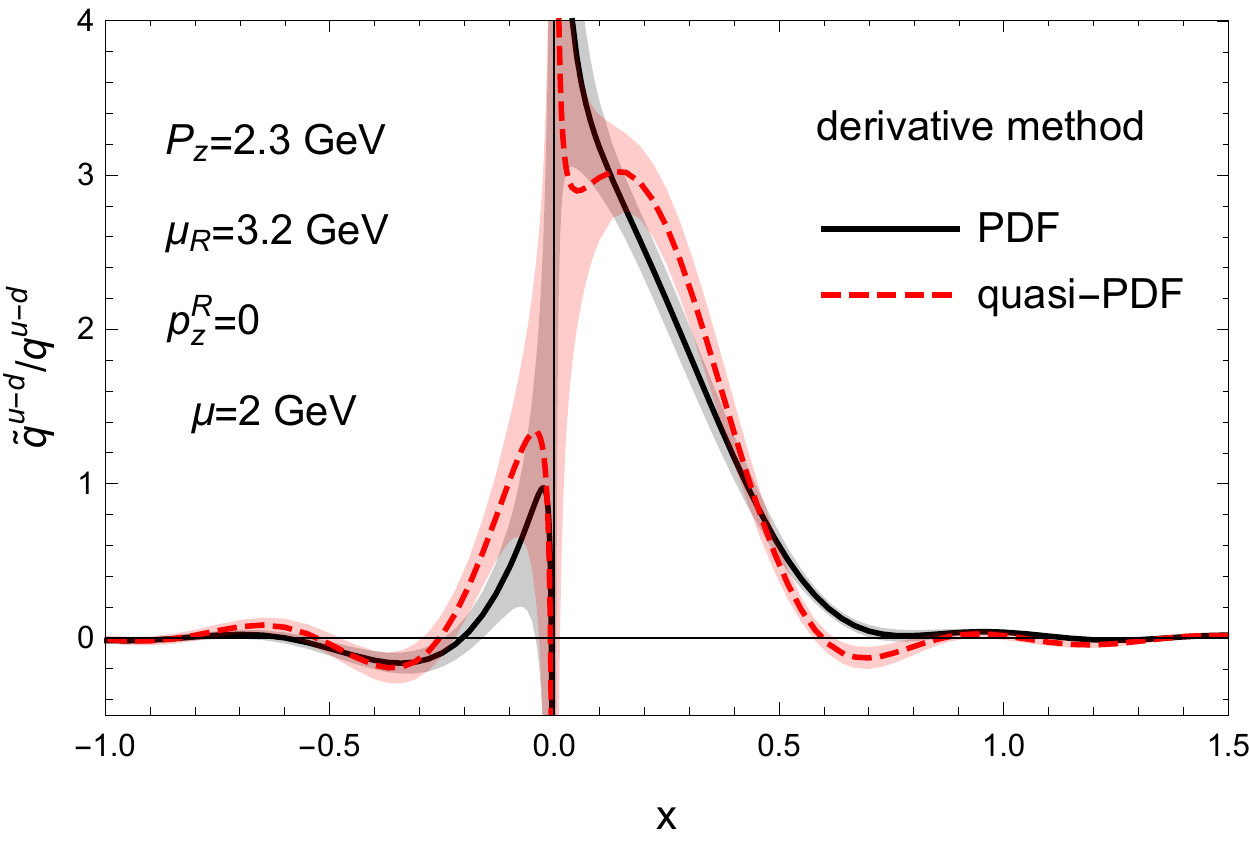}
\caption{\label{fig:matchingPDF} The quasi-PDF (dashed-red) with nucleon boost momentum 2.3 GeV and matched PDF (solid-black) at $\mu=2$~GeV using minimal projection, with the RI/MOM parameters $p_z^R=0$ and $\mu_R=3.2$~GeV. The upper and lower figures are obtained  using the  derivative and cutoff methods to perform  Fourier transformation. The matching strategy has an important impact on the final results for PDFs.}
\end{figure}

In the following, we explain the details to include these systematic uncertainties.

 {1) Fourier transformation.  As shown in Fig.~\ref{fig:renormalized_h-tilde},  the  $\widetilde{h}_R(z)$ with $P_z$=2.3 GeV is consistent with zero when $z>12a$. Thus in the standard matching from quasi-PDF to PDF, it is reasonable to truncate the results at $z=12a$. With this spirit,  the quasi-PDF and matched PDF using the standard FT are shown in the upper panel of Fig.~\ref{fig:matchingPDF}, from which one can see  the matched PDF shows an oscillatory behavior. A ``derivative'' method was proposed in Ref. ~\cite{Lin:2017ani}  to cure  this  oscillatory behavior. To be concrete, one takes the derivative of the renormalized nucleon matrix elements $\partial_z\widetilde{h}_R(z)$, whose Fourier transform differs from the original matrix element in a known way:
\begin{equation}
\label{eq:derivative}
\widetilde{q}_R(x) = \int^\infty_{-\infty} \frac{dz}{2\pi} \frac{ie^{i x P_z z}}{x} \partial_z\widetilde{h}_R(z),
\end{equation}
provided that $\widetilde{h}_R(z)$ goes to zero as $|z|\to\infty$. With the same truncation, the result is shown in the lower panel of Fig.~\ref{fig:matchingPDF} and apparently  the oscillatory behavior is less severe. Besides, results obtained using the derivative method is consistent with the standard FT method in most kinematics region except   at  small $x$. This is anticipated  as two methods only differ at the large $z$ region where we have made the truncation.  We show  the  difference as the dot-dashed-blue line in Fig.~\ref{fig:error}, together with the error from varying the truncation from $z=10a$ to $14a$ (dotted-green line). }

\begin{figure}[tbp]
\includegraphics[width=0.45\textwidth]{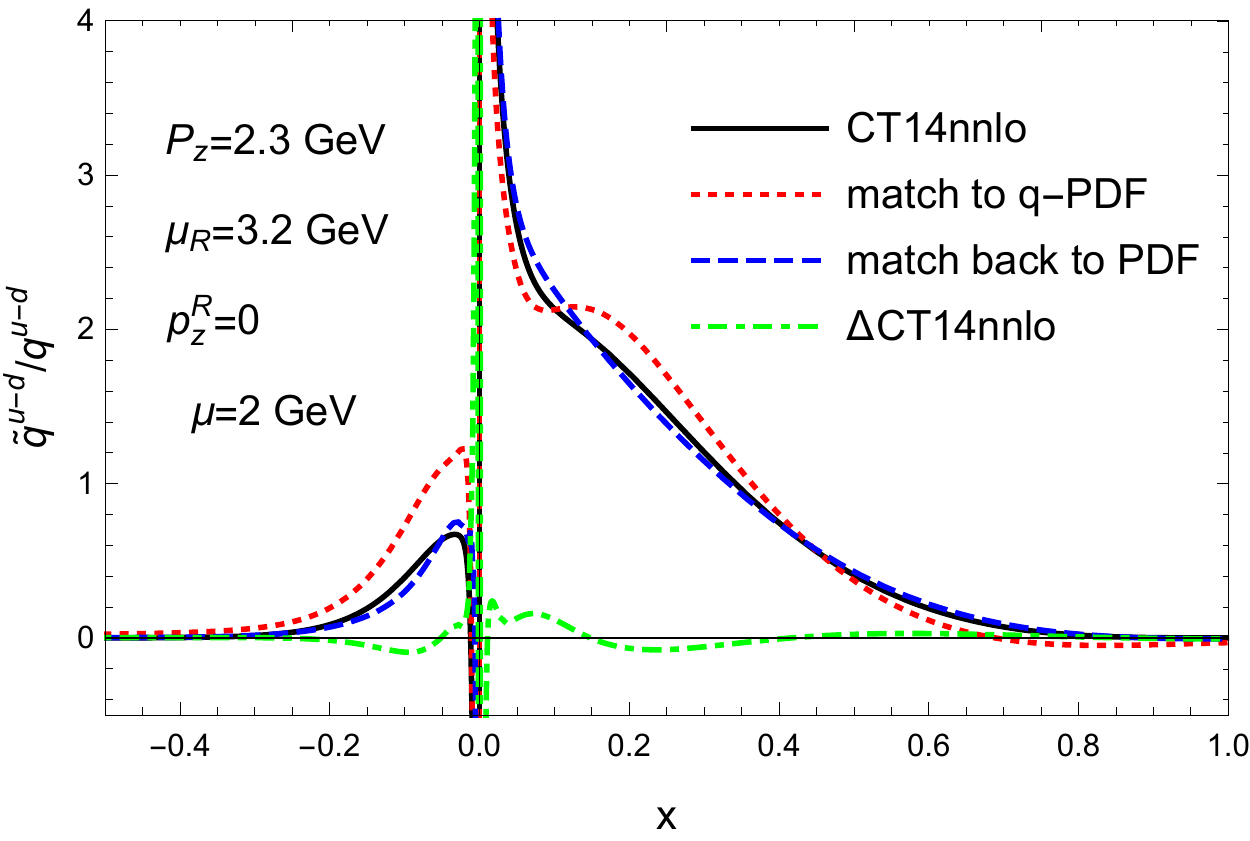}
\includegraphics[width=0.45\textwidth]{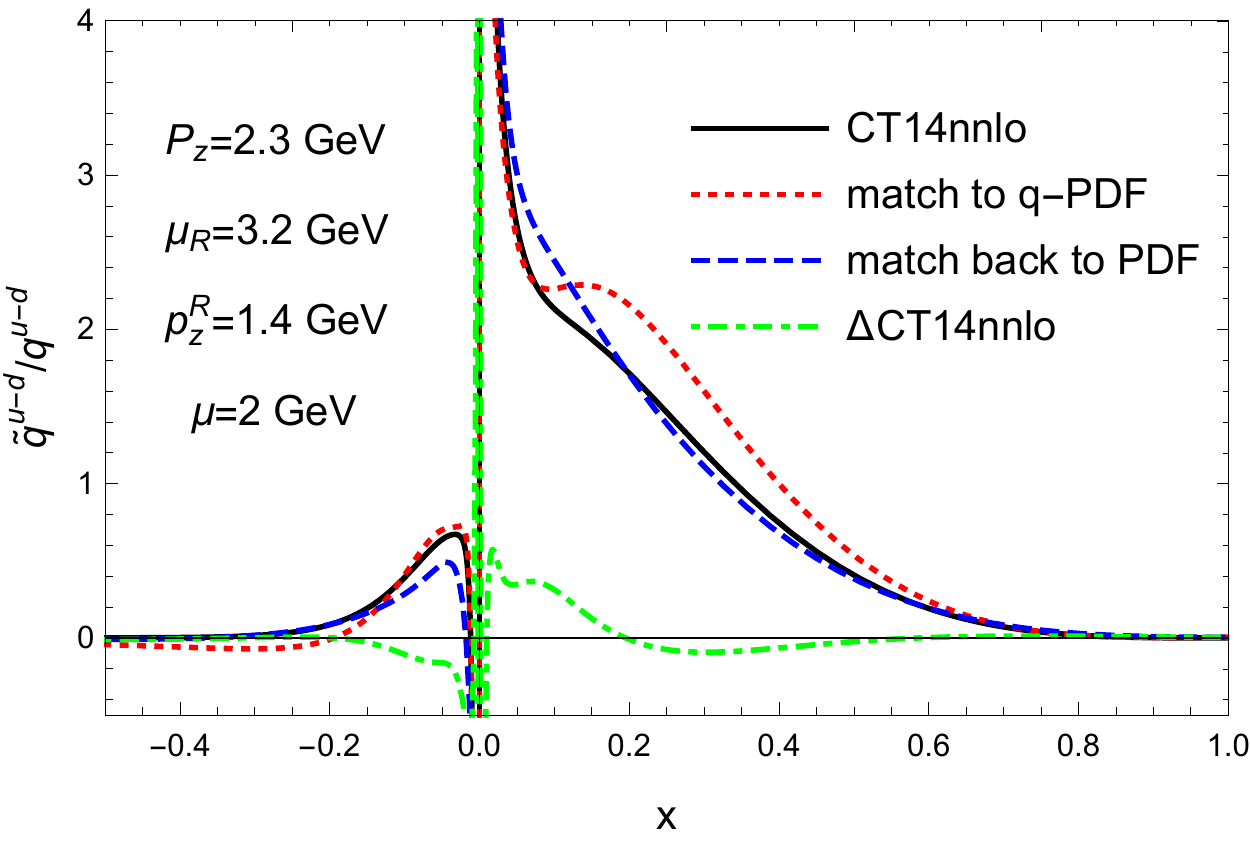}
\caption{\label{fig:mathcing_error_test} Effects of inversion matching formula using minimal projection: The solid-black, dotted-red, dotted-blue, and dot-dashed-green lines represent CT14nnlo PDF, applying inverse matching from CT14nnlo PDF~\cite{Dulat:2015mca} to quasi-PDF, applying matching again to get back to the PDF, the difference between PDF with iterative matching and the original CT14nnlo PDF. The upper (lower) figure corresponds to $p_z^R=0$ (1.4 GeV). These plots show that the method we used to invert the matching formula is less reliable for small $|x|$. The difference shown by the dot-dashed-green curve is taken into account into our systematic error.}
\end{figure}

 {2) Unphysical scales $p_z^R$ and $\mu_R$. There are two unphysical scales $p_z^R$ and $\mu_R$ introduced in RI/MOM.  In principle, when matching the quasi-PDF matrix element onto lightcone PDF, the dependence on these two scales  in the matrix element   should exactly  cancel with that in the matching kernel.  However, since the quasi-PDF matrix element is non-perturbatively renormalized  on the Lattice, while the matching coefficient is  calculated at one-loop order in perturbation theory, there will be residual dependence on these two scales after the perturbative matching. To estimate the residual $p_z^R$ and $\mu_R$ dependence, we choose   {$p_z^R=0$ GeV and $\mu_R=3.2$ GeV} as the central value,  and vary $p_z^R$ from -1.4 to 1.4 GeV (dashed-red line) and $\mu_R$ from 2.4 to 3.9 GeV (thick-solid-orange line). The   difference between  these matched PDFs is treated  as the systematics of the residual dependence on unphysical scales, in Fig.~\ref{fig:error}. As shown in the figure, the systematic uncertainty due to the $\mu_R$ dependence is small compared  to the other sources, but the residual $p_z^R$ dependence could be sizable.}

 {3) Dependencies on the projection. There are two projections discussed in this work:  the minimum and ``$\slashed p$" projections.  With $p_z^R=0$, the projection dependence in both the NPR factor are less than 5\%  for all the $z$, and vanishes in the 1-loop perturbative matching.  Thus one can expect that the difference due to the projections is also small, as depicted  in the long-dashed-magenta lines in Fig.~\ref{fig:error}.}

4) Inversion of matching. To extract the PDF from the quasi-PDF, one needs to invert the factorization formula Eq. (\ref{eq:fact}). This could be done by   changing the sign of $\alpha_s$ in $C$,  and convoluting the new matching coefficient with $\widetilde{q}$. More explicitly, we  have
\begin{align}\label{eq:inverse_fact}
q(x,\mu)=&\int_{-\infty}^\infty {dy\over |y|}\: C'\left({x\over y},r,\frac{yP_z}{\mu},\frac{yP_z}{p_z^R}\right)\widetilde{q}(y,P_z, p^R_z,\mu_R)\nonumber\\
&+\mathcal{O}\left({M^2\over P_z^2},{\Lambda_{\text{QCD}}^2\over x^2 P_z^2},\alpha_s^2\right)\,,
\end{align}
where $C'=C(\alpha_s\to-\alpha_s)$. We estimate the error due to inverting the factorization formula  by  {starting from the PDF from a global analysis~\cite{Dulat:2015mca}, applying Eq.(\ref{eq:fact}) and then Eq. (\ref{eq:inverse_fact}) to return to the PDF. This manipulation should give the same lightcone PDF}. However, since the matching  is   only accurate up to $O(\alpha_s)$, the two results would  differ, and the difference gives a good estimate of the systematic error coming from the inversion and higher order corrections. This is shown in Fig.~\ref{fig:mathcing_error_test}, from which one can find   that the error only  becomes sizable  when $|x|$ is small.  {This is expected because the relevant momentum scale is $xP^z$ such that higher order corrections become large at small $x$}.

There are more sophisticated methods to invert the factorization, such as using a recursion procedure. However, as we can see in Fig.~\ref{fig:error}, the systematic error caused by the matching procedure (thick-dashed-cyan line) is also smaller than those from the first two sources in most regions.

 As shown in Fig.~\ref{fig:error}, one can find  that the dominant   uncertainties arise  from  the $p_z^R$ dependence and  from FT. With $p_z^R=0$, the uncertainties  from different  projections, and  inversion and the matching are typically  less than $10\%$ except in the region with very small $|x|$. The uncertainty  from the $\mu_R$ dependence is even smaller. With  $p_z^R=1.4$ GeV, uncertainties from these  three sources are getting  larger in magnitude, but still smaller than those from the two major sources. 

\section{Final Results for PDF}\label{sec:final_results}

\begin{figure}[tbp]
\includegraphics[width=.48\textwidth]{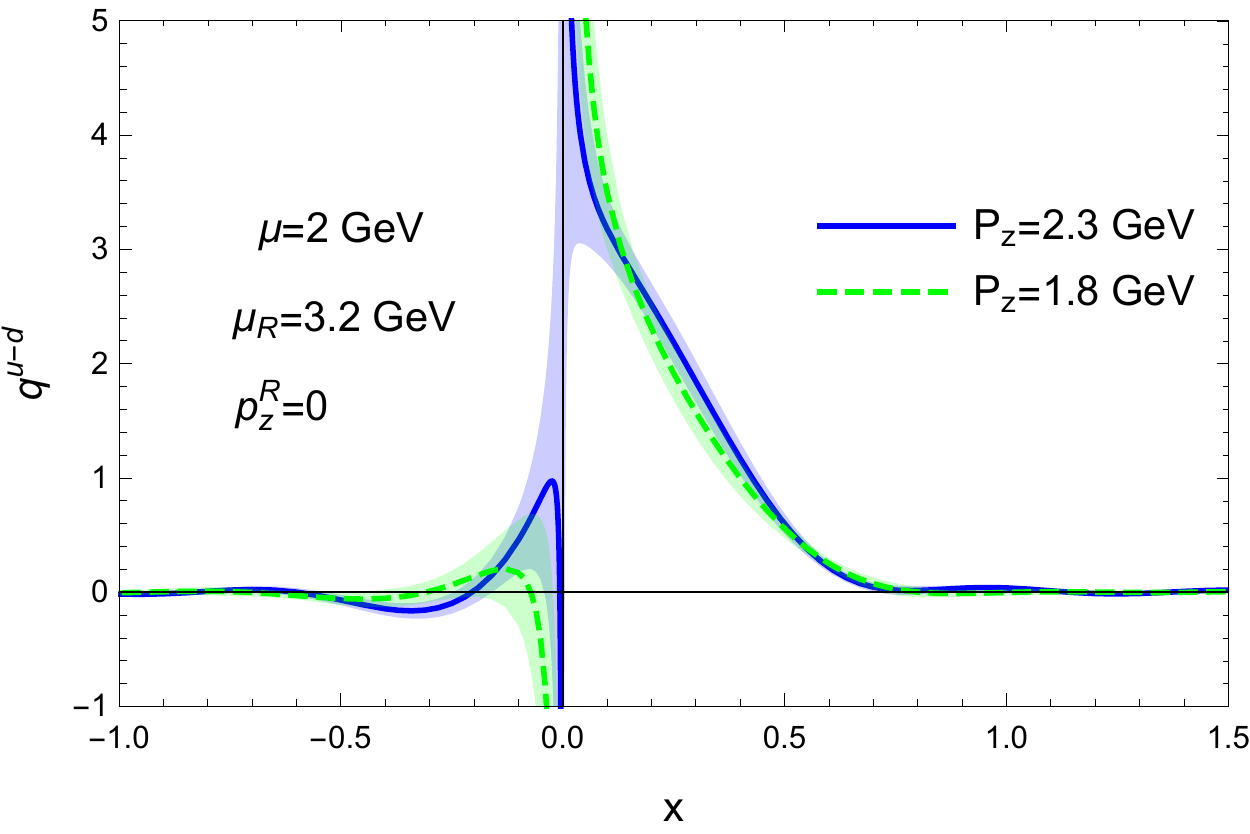}
\caption{\label{fig:matchingPDFallPz} Nucleon boost momentum dependence of the matched unpolarized isovector PDFs: the dotted-green and solid-blue lines correspond to the nucleon momentum $P_z$ to be 1.8 and 2.3 GeV, respectively. The cutoff of Fourier transformation is chosen to be $zP^z\sim 12$ ($z=15a$ for $P_z=1.8$ GeV and $z=12a$ for $P_z=2.3$ GeV).}
\end{figure}

With the derivative method of FT and the matching using $p_z^R=0$ GeV and $\mu_R=3.2$ GeV, we show the dependence on the nucleon boosted momentum in Fig.~\ref{fig:matchingPDFallPz} with the statistical uncertainties.  They are consistent with each other as we can expect from the consistency of the quasi-PDF matrix element results in Fig.~\ref{fig:renormalized_h-tilde}.

\begin{figure}[tbp]
\includegraphics[width=.48\textwidth]{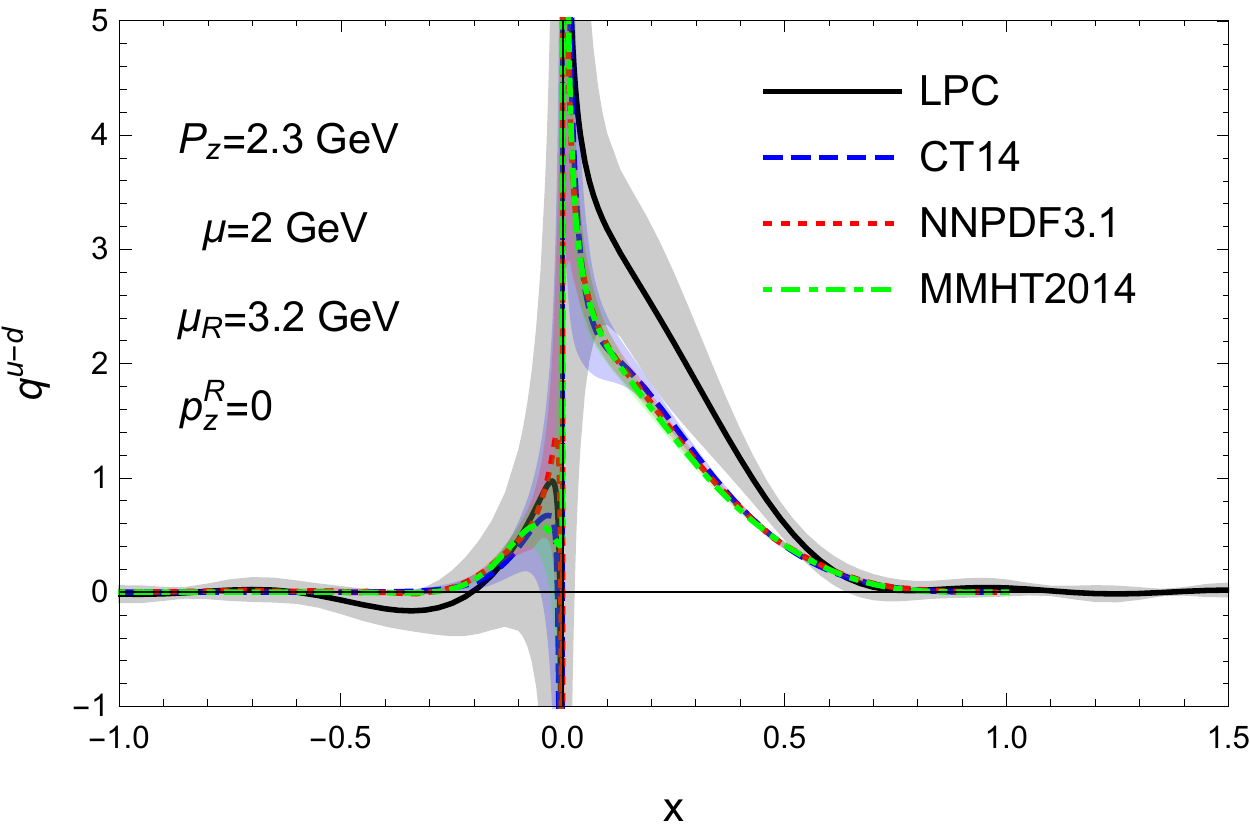}
\caption{\label{fig:finalPDF} Results for PDF at $\mu=2$~GeV calculated from RI/MOM quasi-PDF at nucleon momentum $P_z=2.3$~GeV: Comparing with CT14nnlo (90CL) \cite{Dulat:2015mca}, NNPDF3.1 (68CL) \cite{Ball:2017nwa}, and MMHT2014 (68CL) \cite{Harland-Lang:2014zoa}. Our results agree  with the global-analysis within   uncertainties.}
\end{figure}

 Finally, we show our results for PDF  and a comparison with global-analysis~\cite{Dulat:2015mca,Ball:2017nwa,Harland-Lang:2014zoa} in Fig.~\ref{fig:finalPDF}. As can be seen from the plot, our results show a reasonable agreement in the large-$x$ region, but at small-$x$ region there exists notable difference  majorly due to the systematic uncertainties from the FT truncation method and also the $p_z^R$ dependence.

\section{Summary}\label{sec:summary}

In this paper, we have  studied  the quasi-PDF defined with $\gamma^t$ which is free from mixing at $\mathcal{O}(a^0)$. We  have used $M_\pi=356$~MeV Lattice data to demonstrate the matching procedure and show that the excited state contamination is well under control. The one-loop matching coefficient is calculated and we have  discussed the sources of systematic errors as well as the choice of the projection in detail.

We have found  that the systematic uncertainties from the FT truncation method and also the $p_z^R$ dependence are sizable. But  those uncertainties  from $\mu_R$, inversion of matching and choice of projection are relatively minor with $p_z^R$=0. At the same time, the significant change from quasi-PDF to matched PDF suggests that  higher-loop corrections are needed as exhibited in Fig.~\ref{fig:matchingPDF}.

Controlling   systematic uncertainty from the excited state is very challenging since the relative uncertainty grows very fast when either source-sink separation $t_\text{seq}$ or nucleon momentum $P_z$ become large. The two-state fit with smaller separation provides a possibility to obtain a precise result in small $t_\text{seq}<1$fm  region, while for an accurate measurement at large separation using very high statistics, estimating the systematic uncertainty of such a fit is still needed.

Besides the uncertainties that we have studied, in  the future we plan to investigate other systematics such as Lattice discretization and finite volume effects~\cite{Lin:2019ocg} as well as higher twist contributions that affect the small-$x$ result. The latter can be improved with larger nucleon momentum and estimated by extrapolating to infinite nucleon momentum.

Our final result for lightcone PDF agrees with the  global analysis in the large-$x$ region, which gives an encouraging signal  that   LaMET may allow  us to  precisely access  parton physics in the future.

\section*{Acknowledgment}
We thank the CLS Collaboration for sharing the Lattices used to perform this study. The LQCD calculations were performed using the Chroma software suite~\cite{Edwards:2004sx}.  We thank Xiangdong Ji for useful discussions.  Y.-S. Liu thanks Jun Gao and Shuai Zhao for useful discussions, and Y.-B. Yang thanks Huey-Wen Lin for the discussion on part of the simulation setup.
The numerical calculation is supported by Center for HPC of Shanghai Jiao Tong University, HPC Cluster of ITP-CAS, Jiangsu Key Lab for NSLSCS and the Strategic Priority Research Program of Chinese Academy of Sciences Grant No. XDC01040100.
J.-W. Chen is partly supported by the Ministry of Science and Technology, Taiwan, under Grant No. 105-2112-M-002-017-MY3 and the Kenda Foundation. L.-C. Jin is supported by the Department of Energy, Laboratory Directed Research and Development (LDRD) funding of BNL, under contract DE-EC0012704. Y.-S. Liu is supported by Science and Technology Commission of Shanghai Municipality under Grant No.16DZ2260200 and National Natural Science Foundation of China under Grant No.11905126.
P. Sun is supported by Natural Science Foundation of China under
grant No. 11975127.
W. Wang is supported in part by Natural Science Foundation of China under
grant No. 11575110, 11735010, 11911530088,  by Natural Science Foundation of Shanghai under grant No. 15DZ2272100.
Y.-B. Yang is  partly supported by the US National Science Foundation under grant PHY 1653405 ``CAREER: Constraining Parton Distribution Functions for New-Physics Searches'' and the CAS Pioneer Hundred Talents Program. J.-H. Zhang is supported by Natural Science Foundation of China under
grant No.  11975051, and  the SFB/TRR-55 grant ``Hadron Physics from Lattice QCD''. Q.-A. Zhang is supported by the China Postdoctoral Science Foundation and the National Postdoctoral Program for Innovative Talents (Grant No. BX20190207). Y. Zhao is supported in part by the U.S.~Department of Energy, Office of Science, Office of Nuclear Physics, under grant Contract Number DE-SC0011090, DE-SC0012704 and within the framework of the TMD Topical Collaboration.

\appendix

\begin{widetext}

\section*{Appendix}\label{sec:Appendix}
\subsection{One-loop quasi-PDF with $\gamma^\alpha$ in general covariant gauge}\label{app:one-loop}
The gluon propagator in the general covariant gauge is
\begin{align}
iD_\tau^{\mu\nu}(k)=-\frac{i}{k^2}\left[g^{\mu\nu}-(1-\tau)\frac{k^\mu k^\nu}{k^2}\right]\,.
\end{align}
For general $\Gamma=\gamma^\alpha$, the one-loop result can be expressed as
\begin{align}
\widetilde{q}^{(1)}(x,p,\rho)=\mbox{Tr}\left[\left(\left[\widetilde{f}_\alpha(x,\rho)\right]_+\gamma^\alpha+\left[\widetilde{f}_z(x,\rho)\right]_+\frac{p_\alpha}{p_z}\gamma^z+\left[\widetilde{f}_p(x,\rho)\right]_+\frac{p_\alpha\slashed{p}}{p^2}\right){\cal P}\right]\,,
\end{align}
where
\begin{align}
\widetilde{f}_\alpha(x,\rho)&=\frac{\alpha_s C_F}{2\pi}\left\{
\begin{array}{lc}
\frac{x-\rho}{(1-x)(1-\rho)}+\frac{2x(2-x)-\rho(1+x)}{2(1-x)(1-\rho)^{3/2}}\ln\frac{2x-1+\sqrt{1-\rho}}{2x-1-\sqrt{1-\rho}} & x>1\\
\frac{-3x+2x^2+\rho}{(1-x)(1-\rho)}+\frac{2x(2-x)-\rho(1+x)}{2(1-x)(1-\rho)^{3/2}}\ln\frac{1+\sqrt{1-\rho}}{1-\sqrt{1-\rho}} & 0<x<1\\
-\frac{x-\rho}{(1-x)(1-\rho)}-\frac{2x(2-x)-\rho(1+x)}{2(1-x)(1-\rho)^{3/2}}\ln\frac{2x-1+\sqrt{1-\rho}}{2x-1-\sqrt{1-\rho}} & x<0
\end{array}\right.\nonumber\\
&+\frac{\alpha_s C_F}{2\pi}(1-\tau)\left\{
\begin{array}{lc}
\frac{\rho(-3x+2x^2+\rho)}{2(1-x)(1-\rho)(4x-4x^2-\rho)}+\frac{-\rho}{4(1-\rho)^{3/2}}\ln\frac{2x-1+\sqrt{1-\rho}}{2x-1-\sqrt{1-\rho}} & x>1\\
\frac{-x+\rho}{2(1-x)(1-\rho)}+\frac{-\rho}{4(1-\rho)^{3/2}}\ln\frac{1+\sqrt{1-\rho}}{1-\sqrt{1-\rho}} & 0<x<1\\
-\frac{\rho(-3x+2x^2+\rho)}{2(1-x)(1-\rho)(4x-4x^2-\rho)}-\frac{-\rho}{4(1-\rho)^{3/2}}\ln\frac{2x-1+\sqrt{1-\rho}}{2x-1-\sqrt{1-\rho}} & x<0
\end{array}\right.,
\end{align}
\begin{align}
\widetilde{f}_z(x,\rho)&=\frac{\alpha_s C_F}{2\pi}\left\{
\begin{array}{lc}
\begin{array}{l}
\frac{-2\rho(1-7x+6x^2)-\rho^2(1+2x)}{(1-\rho)^2(4x-4x^2-\rho)}g_{z\alpha}+\frac{4x(1-3x+2x^2)-\rho(2-11x+12x^2-4x^3)-\rho^2}{(1-x)(1-\rho)^2(4x-4x^2-\rho)}\\
\quad+\left[\frac{\rho(4-6x-\rho)}{2(1-\rho)^{5/2}}g_{z\alpha}+\frac{2-4x+4x^2-5x\rho+2x^2\rho+\rho^2}{2(1-x)(1-\rho)^{5/2}}\right]\ln\frac{2x-1+\sqrt{1-\rho}}{2x-1-\sqrt{1-\rho}}
\end{array} & x>1\\
\begin{array}{l}
\frac{-2+2x-\rho(1-4x)}{(1-\rho)^2}g_{z\alpha}+\frac{(-1+2x)(2-3x+\rho)}{(1-x)(1-\rho)^2}\\
\quad+\left[\frac{\rho(4-6x-\rho)}{2(1-\rho)^{5/2}}g_{z\alpha}+\frac{2-4x+4x^2-5x\rho+2x^2\rho+\rho^2}{2(1-x)(1-\rho)^{5/2}}\right]\ln\frac{1+\sqrt{1-\rho}}{1-\sqrt{1-\rho}}
\end{array} & 0<x<1\\
\begin{array}{l}
-\frac{-2\rho(1-7x+6x^2)-\rho^2(1+2x)}{(1-\rho)^2(4x-4x^2-\rho)}g_{z\alpha}-\frac{4x(1-3x+2x^2)-\rho(2-11x+12x^2-4x^3)-\rho^2}{(1-x)(1-\rho)^2(4x-4x^2-\rho)}\\
\quad-\left[\frac{\rho(4-6x-\rho)}{2(1-\rho)^{5/2}}g_{z\alpha}+\frac{2-4x+4x^2-5x\rho+2x^2\rho+\rho^2}{2(1-x)(1-\rho)^{5/2}}\right]\ln\frac{2x-1+\sqrt{1-\rho}}{2x-1-\sqrt{1-\rho}}
\end{array} & x<0
\end{array}\right.\nonumber\\
&+\frac{\alpha_s C_F}{2\pi}(1-\tau)\left\{
\begin{array}{lc}
\begin{array}{l}
\frac{\rho(1-2x)[-4x(1-x)(2+\rho)+3\rho^2]}{2(1-\rho)^2(4x-4x^2-\rho)^2}g_{z\alpha}+\frac{\rho[-4x(2-9x+6x^2)+\rho(1-10x+2\rho)]}{2(1-\rho)^2(4x-4x^2-\rho)^2}\\
\quad+\frac{\rho[(2+\rho)g_{z\alpha}+3)]}{4(1-\rho)^{5/2}}\ln\frac{2x-1+\sqrt{1-\rho}}{2x-1-\sqrt{1-\rho}}
\end{array} & x>1\\
\frac{-3\rho g_{z\alpha}-1-2\rho}{2(1-\rho)^2}+\frac{\rho[(2+\rho)g_{z\alpha}+3)]}{4(1-\rho)^{5/2}}\ln\frac{1+\sqrt{1-\rho}}{1-\sqrt{1-\rho}} & 0<x<1\\
\begin{array}{l}
-\frac{\rho(1-2x)[-4x(1-x)(2+\rho)+3\rho^2]}{2(1-\rho)^2(4x-4x^2-\rho)^2}g_{z\alpha}-\frac{\rho[-4x(2-9x+6x^2)+\rho(1-10x+2\rho)]}{2(1-\rho)^2(4x-4x^2-\rho)^2}\\
\quad-\frac{\rho[(2+\rho)g_{z\alpha}+3)]}{4(1-\rho)^{5/2}}\ln\frac{2x-1+\sqrt{1-\rho}}{2x-1-\sqrt{1-\rho}}
\end{array} & x<0
\end{array}\right.,
\end{align}
\begin{align}
\widetilde{f}_p(x,\rho)&=\frac{\alpha_s C_F}{2\pi}\left\{
\begin{array}{lc}
\begin{array}{l}
\frac{-4x\rho(3-5x+2x^2)+\rho^2(4-3x+4x^2-4x^3)-\rho^3}{(1-x)(1-\rho)^2(4x-4x^2-\rho)}g_{z\alpha}+\frac{-2x\rho(5-6x)+\rho^2(3-2x)}{(1-\rho)^2(4x-4x^2-\rho)}\\
\quad+\left[\frac{-2\rho(1-4x+2x^2)-\rho^2(2-x+2x^2)+\rho^3}{2(1-x)(1-\rho)^{5/2}}g_{z\alpha}+\frac{-\rho(2-6x+\rho)}{2(1-\rho)^{5/2}}\right]\ln\frac{2x-1+\sqrt{1-\rho}}{2x-1-\sqrt{1-\rho}}
\end{array} & x>1\\
\begin{array}{l}
\frac{\rho(1-2x)(4-3x-\rho)}{(1-x)(1-\rho)^2}g_{z\alpha}+\frac{-2x+3\rho-4x\rho}{(1-\rho)^2}\\
\quad+\left[\frac{-\rho(2-8x+4x^2)-\rho^2(2-x+2x^2)+\rho^3}{2(1-x)(1-\rho)^{5/2}}g_{z\alpha}+\frac{-\rho(2-6x+\rho)}{2(1-\rho)^{5/2}}\right]\ln\frac{1+\sqrt{1-\rho}}{1-\sqrt{1-\rho}}
\end{array} & 0<x<1\\
\begin{array}{l}
-\frac{-4x\rho(3-5x+2x^2)+\rho^2(4-3x+4x^2-4x^3)-\rho^3}{(1-x)(1-\rho)^2(4x-4x^2-\rho)}g_{z\alpha}-\frac{-2x\rho(5-6x)+\rho^2(3-2x)}{(1-\rho)^2(4x-4x^2-\rho)}\\
\quad-\left[\frac{-2\rho(1-4x+2x^2)-\rho^2(2-x+2x^2)+\rho^3}{2(1-x)(1-\rho)^{5/2}}g_{z\alpha}+\frac{-\rho(2-6x+\rho)}{2(1-\rho)^{5/2}}\right]\ln\frac{2x-1+\sqrt{1-\rho}}{2x-1-\sqrt{1-\rho}}
\end{array} & x<0
\end{array}\right.\nonumber\\
&+\frac{\alpha_s C_F}{2\pi}(1-\tau)\left\{
\begin{array}{lc}
\begin{array}{l}
\frac{16x\rho(1-3x+2x^2)+4x^2\rho^2(3-2x)-\rho^3(5-2x)+2\rho^4}{2(1-\rho)^2(4x-4x^2-\rho)^2}g_{z\alpha}\\
\quad+\frac{\rho(1-2x)[16x(1-x)-2\rho(1+2x-2x^2)-\rho^2]}{2(1-\rho)^2(4x-4x^2-\rho)^2}+\frac{-\rho(4-\rho)(g_{z\alpha}+1)}{4(1-\rho)^{5/2}}\ln\frac{2x-1+\sqrt{1-\rho}}{2x-1-\sqrt{1-\rho}}
\end{array} & x>1\\
\frac{\rho(5-2\rho)g_{z\alpha}+2+\rho}{2(1-\rho)^2}+\frac{-\rho(4-\rho)(g_{z\alpha}+1)}{4(1-\rho)^{5/2}}\ln\frac{1+\sqrt{1-\rho}}{1-\sqrt{1-\rho}} & 0<x<1\\
\begin{array}{l}
-\frac{16x\rho(1-3x+2x^2)+4x^2\rho^2(3-2x)-\rho^3(5-2x)+2\rho^4}{2(1-\rho)^2(4x-4x^2-\rho)^2}g_{z\alpha}\\
\quad-\frac{\rho(1-2x)[16x(1-x)-2\rho(1+2x-2x^2)-\rho^2]}{2(1-\rho)^2(4x-4x^2-\rho)^2}-\frac{-\rho(4-\rho)(g_{z\alpha}+1)}{4(1-\rho)^{5/2}}\ln\frac{2x-1+\sqrt{1-\rho}}{2x-1-\sqrt{1-\rho}}
\end{array} & x<0
\end{array}\right. .
\end{align}

\subsection{One-loop quasi-PDF with $\Gamma=\gamma^z$ in Landau gauge}\label{app:one-loop_gamma_z}
For $\Gamma=\gamma^z$, we can obtain the matching coefficient Eq. (\ref{eq:matching_coeff}) using the general formula with similar definition of the minimal and $\slashed p$ projections in Sec. \ref{sec:one_loop_matching}. The bare matching coefficients are
\begin{align}
f_{1,mp}\left(x,\frac{p_z}{\mu}\right)=f_{1,\slashed p}\left(x,\frac{p_z}{\mu}\right)=\frac{\alpha_s C_F}{2\pi}\left\{
\begin{array}{lc}
\displaystyle \frac{1+x^2}{1-x}\ln\frac{x}{x-1}+1 & x>1\\
\displaystyle \frac{1+x^2}{1-x}\ln\frac{4x(1-x)p_z^2}{\mu^2}+\frac{2-5x+x^2}{1-x} & 0<x<1\\
\displaystyle -\frac{1+x^2}{1-x}\ln\frac{x}{x-1}-1 & x<0
\end{array} \right. ,
\end{align}
and the corresponding counterterms are
\begin{align}
f_{2,mp}(x,r)=\frac{\alpha_s C_F}{2\pi}\left\{
\begin{array}{lc}
\frac{3r-(1-2x)^2}{2(r-1)(1-x)}-\frac{4x^2(2-3r+2x+4rx-12x^2+8x^3)}{(r-1)(r-4x+4x^2)^2}+\frac{2-3r+2x^2}{(r-1)^{3/2}(x-1)}\tan^{-1}\frac{\sqrt{r-1}}{2x-1} & x>1\\
\frac{1-3r+4x^2}{2(r-1)(1-x)}+\frac{-2+3r-2x^2}{(r-1)^{3/2}(1-x)}\tan^{-1}\sqrt{r-1} & 0<x<1\\
-\frac{3r-(1-2x)^2}{2(r-1)(1-x)}+\frac{4x^2(2-3r+2x+4rx-12x^2+8x^3)}{(r-1)(r-4x+4x^2)^2}-\frac{2-3r+2x^2}{(r-1)^{3/2}(x-1)}\tan^{-1}\frac{\sqrt{r-1}}{2x-1} & x<0
\end{array} \right.
\end{align}
\begin{align}
f_{2,\slashed{p}}(x,r)=\frac{\alpha_s C_F}{2\pi}\left\{
\begin{array}{lc}
1+\frac{r}{2}\frac{r(3-4x)-8x(x-1)^2}{(x-1)(r-4x+4x^2)^2}+\frac{-2+r-2x^2}{\sqrt{r-1}(x-1)}\tan^{-1}\frac{\sqrt{r-1}}{2x-1} & x>1\\
\frac{1-6x}{2(1-x)}+\frac{2-r+2x^2}{\sqrt{r-1}(1-x)}\tan^{-1}\sqrt{r-1} & 0<x<1\\
-1-\frac{r}{2}\frac{r(3-4x)-8x(x-1)^2}{(x-1)(r-4x+4x^2)^2}-\frac{-2+r-2x^2}{\sqrt{r-1}(x-1)}\tan^{-1}\frac{\sqrt{r-1}}{2x-1} & x<0
\end{array} \right.\,.
\end{align}
The result was also calculated in Ref.~\cite{Stewart:2017tvs}. Although the matching coefficient with $\Gamma=\gamma^z$ is not useful for isovector unpolarized PDF because it suffers from operator mixing in renormalization procedure, it can be used for isovector helicity PDF due to different symmetry properties.

\end{widetext}

%\bibliographystyle{apsrev4-1}
%\bibliography{bibliography}

\end{document}